\documentclass{aa}  

\usepackage{graphicx}
\usepackage{txfonts}
\usepackage{hyperref}
\usepackage{xcolor} 
\usepackage{orcidlink}
\begin{document}

\title{A new magnitude--redshift relation based on Type~Ia supernovae}

\authorrunning{Rodr\'iguez, \'O. and Clocchiatti, A.}
\titlerunning{A new magnitude--redshift relation based on SNe Ia}

\author{
\'Osmar Rodr\'iguez\inst{1,2}\fnmsep\thanks{Corresponding author: olrodrig@gmail.com}\orcidlink{0000-0001-8651-8772}
\and
Alejandro Clocchiatti\inst{1,2}\orcidlink{0000-0003-3068-4258}
}

\institute{
Pontificia Universidad Cat\'olica de Chile, Vicu\~na Mackenna 4860, Macul, Santiago, Chile
\and
Instituto Milenio de Astrof\'isica (MAS), Nuncio Monse\~nor S\'otero Sanz 100, Of. 104, Santiago, Chile
}

\date{Received X; accepted Y}
 
\abstract{We present a new empirical relation between the standardized magnitude ($m$) of Type Ia supernovae (SNe~Ia) and redshift ($z$). Using Pantheon+ and DES-SN5YR, we find a negative linear correlation between $m-5\log(z(1+z))$ and $z$, implying that their magnitude--redshift relation can be parametrized with just two parameters: an intercept $\mathcal{M}$ and a slope $b$. This relation corresponds to the luminosity distance $d_L(z)=c\,H_0^{-1}z(1+z)10^{bz/5}$ and is valid up to at least $z\simeq1.1$. It outperforms the $\Lambda$CDM and flat $w$CDM models and the (2,1) Pad\'e approximant for $d_L(z)$, and performs comparably to the flat $\Lambda$CDM model and the (2,1) Pad\'e($j_0=1$) model of Hu et al. Furthermore, the relation is relatively stable in the absence of low-$z$ SNe, making it suitable for fitting Hubble diagrams of SNe~Ia without the need to add a low-$z$ sample. In deep fields in particular, assuming that the large-scale density is independent of the comoving radial coordinate, $b\propto q_0+1$. We fit the empirical relation to Hubble diagrams of eight deep-field regions, finding no evidence for anisotropy. The inferred $q_0$ values, ranging from $-0.6$ to $-0.4$, are consistent within $1.6\,\sigma$ and significantly lower than zero, indicating statistically consistent cosmic acceleration across all eight regions. We apply the empirical relation to the DES-Dovekie and Amalgame SN samples, finding $b$ values consistent with those from DES-SN5YR and Pantheon+. Finally, using the empirical relation in the hemispheric comparison method applied to Pantheon+ up to $z=1.1$, we find no evidence for anisotropies in $\mathcal{M}$ and $b$.
}

\keywords{supernovae: general -- cosmological parameters -- cosmology: theory}
\maketitle
%

\section{Introduction}
General Relativity and the assumption of homogeneity and isotropy of the Universe at large scales are two of the foundations of modern cosmology. The latter leads to the Friedmann--Lema\^itre--Robertson--Walker (FLRW) metric, under which Einstein's field equations reduce to the Friedmann equations. 
The first, combined with the FLRW metric, provides a formula for the luminosity distance--redshift relation, $d_L(z)$, in terms of three parameters: the Hubble constant ($H_0$), the matter density ($\Omega_M$), and the cosmological constant ($\Lambda$) with density parameter $\Omega_\Lambda$.

Using Hubble diagrams of Type Ia supernovae (SNe~Ia), \citet{1998AJ....116.1009R} and \citet{1999ApJ...517..565P} found $\Omega_\Lambda>0$ and a deceleration parameter $q_0<0$, indicating that the expansion of the Universe is currently accelerating. In this context, $\Lambda$ is interpreted as a scalar related to a hypothetical dark energy that drives cosmic acceleration. Given that models with $\Lambda$, when fitted to various observational data, yield $\Omega_\Lambda>0$ \citep{2013PhR...530...87W}, the existence of dark energy has been widely accepted, consolidating the $\Lambda$CDM model as the standard model.

Despite its ability to fit observations, the $\Lambda$CDM model has problems of fine-tuning, cosmic coincidence, and tensions in cosmological parameters measured with independent experiments \citep{2022NewAR..9501659P,2025RSPTA.38340022E}. These issues have motivated alternative gravitational theories to explain cosmic acceleration without $\Lambda$ \citep{2016RPPh...79d6902K,2025EPJC...85..298O}. Moreover, cosmic acceleration may be partly an apparent effect associated with the assumption of homogeneity and isotropy and the corresponding use of the FLRW metric \citep{2006JCAP...11..003R,2008GReGr..40..451E,2009PhRvD..80l3512W}.

Cosmography provides a model-independent framework to study cosmic acceleration, expressing $d_L(z)$ in terms of spatial curvature and kinematic parameters, such as $H_0$, $q_0$, and the jerk parameter ($j_0$). This approach, however, relies on the assumption of a homogeneous and isotropic Universe \citep{2022AA...661A..71H}.

Among current SN~Ia samples, such as Pantheon+ \citep{2022ApJ...938..110B} and DES-SN5YR \citep{2024ApJ...975....5S}, a significant fraction of the SNe originates from deep-field surveys. Analyzing deep fields separately allows one to infer direction-dependent parameters, without requiring global homogeneity and isotropy. In particular, comparing deep fields across different directions provides a test of isotropy.

Since angular variations in deep fields are expected to be negligible, and assuming that the large-scale density within each field is independent of the comoving radial coordinate, the metric can be approximated as FLRW on a field-by-field basis. Under this assumption, it might be possible to measure $q_0$ for each deep field using the FLRW metric, without assuming global homogeneity and isotropy. The main limitation of this approach is the small number of low-$z$ SNe~Ia ($z<0.1$) in deep fields, which are crucial for breaking parameter degeneracies \citep{2006PhRvD..74j3518L}.

One approach to reducing the impact of the lack of low-$z$ SNe on the estimation of $q_0$ is to use a $d_L(z)$ relation with as few parameters as possible. In the flat $\Lambda$CDM model, $d_L(z)$ depends only on $H_0$ and $\Omega_M$. The latter, which is used to measure $q_0$, remains relatively stable in the absence of low-$z$ SNe \citep{2022ApJ...938..110B}. To date, the only other $d_L(z)$ relation that accurately fits SN observations with just two parameters is the one proposed by \citet{2024AA...689A.215H}. It corresponds to a third-order Pad\'e approximation with $H_0$ and $q_0$ as free parameters, while $j_0$ is fixed to one, as predicted by the flat $\Lambda$CDM model \citep{2015ApJ...814....7B}.

In this work, we introduce an empirical magnitude--redshift relation ($m(z)$) and the corresponding $d_L(z)$ relation, which provides a straightforward quantitative tool to fit Hubble diagrams of SNe~Ia up to at least $z\simeq1.1$, using only two parameters and without assuming any theoretical model or spatial curvature. This relation is relatively stable in the absence of low-$z$ SNe, making it suitable for studying SN deep fields without the need to add a low-$z$ sample.

The paper is organized as follows. In Sect.~\ref{sec:data_sample}, we describe the SN samples and define the deep-field regions. In Sect.~\ref{sec:methodology}, we present the $m(z)$ relations employed to fit Hubble diagrams. In Sect.~\ref{sec:results}, we present the results, including the new empirical relation, its comparison with five $m(z)$ relations, and its application to the deep-field regions and to other samples. In Sect.~\ref{sec:discussion}, we discuss the derived $j_0$ values, the search for cosmic anisotropies, and future analyses. Our conclusions are summarized in Sect.\ref{sec:conclusions}.

\section{Data samples}\label{sec:data_sample}
Pantheon+\footnote{\url{https://github.com/PantheonPlusSH0ES/DataRelease}} contains data from 1550 SNe~Ia, drawn from 18 surveys, with $0.001\leq z\leq2.261$. DES-SN5YR\footnote{\url{https://doi.org/10.5281/zenodo.12720778}} contains data from 1635 photometrically classified SNe~Ia from the DES-SN program with $0.060\leq z\leq1.121$, and 194 SNe~Ia from four surveys, with $0.025< z<0.093$, referred to as the low-$z$ sample.

Pantheon+ and DES-SN5YR provide RA and Dec coordinates; standardized SN magnitudes ($m_B^\mathrm{corr}$, hereafter $m$), derived from SALT2 (Pantheon+) and SALT3 (DES-SN5YR) light-curve fits, and corrected for stretch, color, host-galaxy mass, and selection bias; redshifts corrected for the CMB dipole and peculiar velocities ($z_\mathrm{HD}$, hereafter $z$); and the covariance matrix ($\mathbf{C}$) to account for statistical and systematic errors.

As done by \citet{2022ApJ...938..110B}, from Pantheon+ we select SNe with $z>0.01$ to minimize the impact of peculiar velocities on the derived cosmological parameters. In the following, we refer to this subsample simply as Pantheon+. Because DES-SN5YR introduced several improvements compared to Pantheon+, their $m$ values are not on the same scale. In particular, there is a constant offset of 0.04\,mag between selection bias corrections in DES-SN5YR and Pantheon+, which does not impact the cosmological results from each separate analysis \citep{2025MNRAS.541.2585V}. Hence, we analyze both samples separately.

\begin{figure*}
\includegraphics[width=0.99\columnwidth]{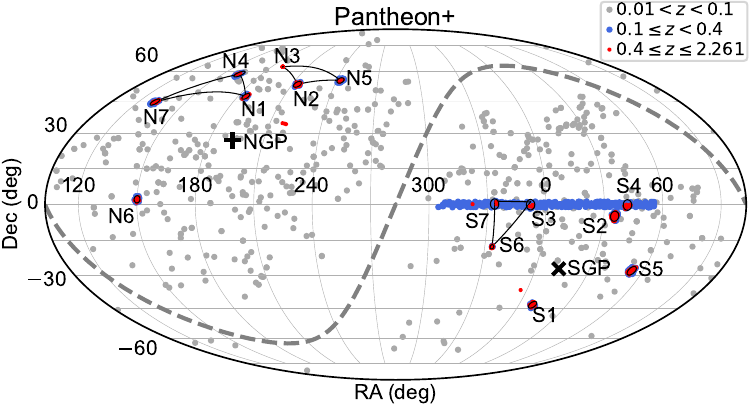}
\includegraphics[width=0.99\columnwidth]{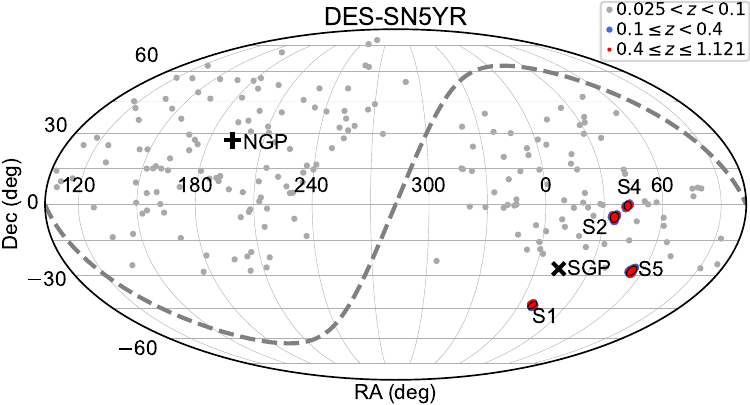}
\caption{Sky distribution of Pantheon+ and DES-SN5YR SNe. Dashed lines indicate the Galactic equator. Plus (cross) symbols mark the north (south) Galactic pole. Circular and triangular regions are also shown.}
\label{fig:SN_sky_distribution}
\end{figure*}

Fig.~\ref{fig:SN_sky_distribution} shows the sky distribution of the SNe in Pantheon+ and DES-SN5YR. SNe with $z\geq0.4$ are concentrated in small regions of the sky, which reflects the location of the SN deep fields. We define 14 circular regions for Pantheon+ and four for DES-SN5YR, referred to as XY, which contain virtually all SNe with $z\geq0.4$. In this notation, X denotes the closest Galactic pole and Y indicates the rank of proximity to that pole. Table~\ref{table:SN_regions} lists the regions, their centers in (RA, Dec), angular radii ($\theta$), number of SNe within each region, and the corresponding $z$ ranges.

\begin{table}
  \centering
  \caption{SN regions}
  \label{table:SN_regions}
  \begin{tabular}{ c r r c c c}
    \hline\hline
    Region & RA (\degr)    & Dec (\degr)     & $\theta$ (\degr)& $z$ range & \#SNe  \\
    \hline
    \multicolumn{6}{c}{Pantheon+} \\
    \hline
 S1   &   8.57 & $-43.36$ & 1.60 & 0.148--0.609 &   34 \\
 S2   &  35.56 & $ -4.88$ & 2.12 & 0.017--1.912 &  134 \\
 S3   & 352.15 & $ -0.09$ & 2.17 & 0.079--0.508 &   34 \\
 S4   &  41.89 & $ -0.21$ & 2.19 & 0.134--0.638 &   57 \\
 S5   &  53.78 & $-28.00$ & 1.89 & 0.103--1.549 &   97 \\
 S6   & 333.83 & $-17.70$ & 0.61 & 0.371--0.789 &   44 \\
 S7   & 333.68 & $  0.20$ & 2.14 & 0.040--0.419 &   43 \\
 N1   & 185.14 & $ 47.10$ & 1.37 & 0.025--0.545 &   24 \\
 N2   & 213.64 & $ 53.01$ & 1.47 & 0.082--1.615 &   64 \\
 N3   & 189.24 & $ 62.22$ & 0.10 & 0.840--2.261 &   13 \\
 N4   & 162.85 & $ 58.30$ & 1.52 & 0.023--0.503 &   20 \\
 N5   & 242.82 & $ 54.96$ & 1.40 & 0.122--0.576 &   30 \\
 N6   & 150.05 & $  2.18$ & 1.51 & 0.047--1.543 &   77 \\
 N7   & 130.55 & $ 44.39$ & 1.44 & 0.071--0.578 &   37 \\ \hline
 S367 & 339.96 & $ -5.10$ & --   & 0.015--0.789 &  163 \\
 N147 & 158.83 & $ 52.31$ & --   & 0.023--0.578 &   89 \\
 N235 & 216.91 & $ 58.56$ & --   & 0.082--2.261 &  107 \\
    \hline
    \multicolumn{6}{c}{DES-SN5YR} \\ 
    \hline
 S1   &   8.63 & $-43.53$ & 1.76 & 0.073--0.818 &  295 \\
 S2   &  35.57 & $ -5.32$ & 2.11 & 0.138--1.044 &  518 \\
 S4   &  42.00 & $ -0.51$ & 1.83 & 0.094--0.733 &  231 \\
 S5   &  53.83 & $-28.09$ & 2.11 & 0.060--1.121 &  591 \\
    \hline
  \end{tabular}
\end{table}

The S1, S2, S4, and S5 regions in Pantheon+ closely match those in DES-SN5YR. In these regions, DES-SN5YR contains 170--490 more SNe than Pantheon+, while 47--100\% of the SNe in the Pantheon+ S1, S2, S4, and S5 regions come from the DES-SN program and are already included in DES-SN5YR. Therefore, we use only DES-SN5YR data for these regions.

The remaining ten Pantheon+ regions contain 13--77 SNe each. To increase the sample while minimizing sky area, we define the triangular N147, N235, and S367 regions. N147 includes N1, N4, N7, and the SNe located within the triangle formed by their centers; similarly for N235 and S367. For S367, we increase the declination of the centers of S3 and S7 by 1.2\degr{} to include more SNe without significantly increasing the area. The triangular regions are shown in Fig.~\ref{fig:SN_sky_distribution}, while their barycenters, $N$, and $z$ ranges are listed in Table~\ref{table:SN_regions}. We refer to the S1, S2, S4, S5, N147, N235, S367, and N6 regions as the deep-field regions.

\section{Methods}\label{sec:methodology}
\subsection{Magnitude--redshift relations}
The $m(z)$ relation for a standard candle with absolute magnitude $M$ is given by
\begin{equation}
m(z) = \mathcal{M}+5\log{\mathcal{D}_L(z)},
\end{equation}
where $\mathcal{M}=M+5\log(c\,H_0^{-1}\,\mathrm{Mpc}^{-1})+ 25$ and $\mathcal{D}_L(z)=d_L(z)H_0/c$. To derive $d_L(z)$ using either a gravitational theory or a cosmographic expansion, one needs the metric of the expanding Universe. The simplest choice is the FLRW metric, characterized by the scale factor $a$ and the curvature parameter $k$. For this metric,
\begin{equation}\label{eq:dL_dM}
d_L(z)=(1+z)d_M(z),
\end{equation}
where $d_M(z)$ is the transverse comoving distance. Defining $H(z)=\dot{a}/a$ and $\Omega_k=-kc^2/H_0^2$, $d_M(z)$ is given by
\begin{equation}
d_M(z)=\frac{c}{H_0}\begin{cases}
\Omega_k^{-1/2}\sinh(\Omega_k^{1/2}\int_0^{z}\frac{H_0}{H(z')}dz')&\Omega_k>0\\
\int_0^{z}\frac{H_0}{H(z')}dz'& \Omega_k=0\\
|\Omega_k|^{-1/2}\sin(|\Omega_k|^{1/2}\int_0^{z}\frac{H_0}{H(z')}dz')&\Omega_k<0
\end{cases}.
\end{equation}

\subsubsection{$\Lambda$CDM, flat $\Lambda$CDM, and flat $w$CDM models}
In the phenomenological generalization of the $\Lambda$CDM model,
\begin{equation}\label{eq:Hz}
H(z)=H_0\sqrt{\Omega_M(1+z)^3+\Omega_k(1+z)^2+\Omega_\Lambda(1+z)^{3(1+w)}},
\end{equation}
where $\Omega_M+\Omega_\Lambda+\Omega_k=1$ and $w$ is the dark energy equation-of-state parameter. We use the $m(z)$ relations for $\Lambda$CDM ($w=-1$), flat $\Lambda$CDM ($\Omega_k=0$, $w=-1$), and the phenomenological flat $w$CDM ($\Omega_k=0$) models.

\subsubsection{Pad\'e cosmography and flat Pad\'e model}
The Pad\'e approximation is a method to approximate a function $F(z)$ as the ratio of two polynomials of orders $A$ and $B$. The rational function is known as the ($A$,$B$) Pad\'e approximant, and its coefficients are expressed in terms of those of the Taylor series of $F(z)$ truncated at order $A+B$ \citep{Baker1996}.

\citet{2022AA...661A..71H}, using Pantheon \citep{2018ApJ...859..101S}, found that the (2,1) Pad\'e approximant for $d_L(z)$ performs better than cosmographic expansions based on Taylor series. More recently, \citet{2024AA...689A.215H} analyzed Pantheon+ and found that this approximant outperforms other Pad\'e forms. We therefore adopt it as the best cosmographic approach.

The (2,1) Pad\'e approximant for $d_L(z)$ used by \citet{2024AA...689A.215H} was derived by \citet{2020MNRAS.494.2576C} for $\Omega_k=0$. To compute the expression valid for any $\Omega_k$, we use the third-order Taylor series for $d_L(z)$ (Eq.~2 of \citealt{2007CQGra..24.5985C}) and the procedure of \citet{Baker1996}, obtaining
\begin{equation}\label{eq:Pade21}
d_L(z)=\frac{cz}{H_0}\frac{6(1-q_0) + (5-8q_0-3q_0^2+2\hat{j_0})z}{6(1-q_0)+2(1-q_0-3q_0^2+\hat{j_0})z}.
\end{equation}
Here, $\hat{j_0}\equiv j_0-\Omega_k$, and 
\begin{equation}\label{eq:qj}
q_0=\left.\frac{\mathrm{d}\ln H(z)}{\mathrm{d}z}\right|_{z=0}-1,\quad
j_0=\frac1{H_0}\left.\frac{\mathrm{d}^2H(z)}{\mathrm{d}z^2}\right|_{z=0}+q_0^2. 
\end{equation}
For $\Omega_k=0$, Eq.~(\ref{eq:Pade21}) is equivalent to that given in \citet{2024AA...689A.215H}. We refer to Eq.~(\ref{eq:Pade21}) as the Pad\'e cosmography.

\citet{2024AA...689A.215H} also found that the (2,1) Pad\'e approximant for $d_L(z)$ with $j_0=1$ fixed performs better than the Pad\'e cosmography, the flat $\Lambda$CDM, $\Lambda$CDM, and flat $w$CDM models. Since $j_0$ was fixed to unity because this is the value for the flat $\Lambda$CDM model, the $d_L(z)$ equation proposed by \citet{2024AA...689A.215H} is not cosmographic, but rather model-based. We refer to Eq.~(\ref{eq:Pade21}) assuming $\hat{j_0}=1$ as the flat Pad\'e model.

\subsubsection{Empirical approach}
To fit the Hubble diagram, we introduce the ansatz
\begin{equation}\label{eq:dL_ansatz}
d_L(z)= \frac{cz}{H_0}(1+z)10^{f(z)/5},
\end{equation}
where the factor $1+z$ is inspired by Eq.~(\ref{eq:dL_dM}) and $f(z)$ is a free function that vanishes as $z\to0$, ensuring $d_L(z)= cz/H_0$ at low $z$. The corresponding $m(z)$ relation is
\begin{equation}\label{eq:mz_ansatz}
m(z) = \mathcal{M} +5\log(z(1+z))+ f(z),
\end{equation}
where the dependence of $f(z)$ on $z$ can be determined empirically from the correlation between $m-5\log(z(1+z))$ and $z$.

\subsection{Parameter estimation and model selection}
Let $\mathbf{v}$ be the vector with the $n_p$ free parameters of $m(z)$. The posterior probability of $m(z)$ is $P \propto p(\mathbf{v})\mathcal{L}$, where $p(\mathbf{v})$ is the prior and $\mathcal{L}=e^{-\chi^2/2}$. Following \citet{2011ApJS..192....1C},
\begin{equation}
\chi^2=\Delta \mathbf{m}^T\mathbf{C}^{-1}\Delta \mathbf{m},
\end{equation}
where $\Delta \mathbf{m}$ is the vector of residuals, whose $i$-th component is
\begin{equation}\label{eq:delta_m}
\Delta m_i=m_i-m(z_i,\mathbf{v}).
\end{equation}
We assume uninformative priors; in this case, the best-fit parameter vector $\mathbf{v}_\mathrm{best}$ that maximizes $P$ is obtained by minimizing $\chi^2$. 

To calculate uncertainties, we use \textsc{emcee} \citep{2013PASP..125..306F}, which samples $P$ using a Markov Chain Monte Carlo (MCMC) process. First, we define the priors as flat distributions: $\mathcal{M}\in(23, 25)$, $\Omega_M\in(0, 1)$, $\Omega_\Lambda\in(0, 1)$, $w\in(-2,0)$, $q_0\in(-2, 1)$, and $\hat{j_0}\in(-4, 7)$. These priors are wide enough to consider them as uninformative. Next, we initialize $10n_p$ walkers in a tiny Gaussian ball around $\mathbf{v}_\mathrm{best}$ and run $10^5$ steps ($10^6$ for the flat $w$CDM model). Then, given an integrated autocorrelation time ($\tau$) provided by \textsc{emcee}, we discard the initial $3\tau$ steps as burn-in and thin by $\tau/2$. With this configuration, we obtain for all parameters a \citet{1992StaSc...7..457G} statistic of $\hat{R} \approx 1.0$, indicating convergence of the MCMC chains, and an effective sample size $n_\text{eff}>2000$ ($n_\text{eff}>400$ for the flat $w$CDM model). Finally, for each parameter, we adopt the 68.27\% confidence interval of its marginalized distribution as the $1\sigma$ uncertainty.

To identify which model of a set of candidates best describes the Hubble diagram, we use the corrected Akaike information criterion (AICc; \citealt{Sugiura1978}) and the Bayesian information criterion (BIC; \citealt{1978AnSta...6..461S}). For each model, we compute $\mathrm{AICc}=\chi^2_\mathrm{min}+2n_pN/(N-n_p-1)$ and $\mathrm{BIC}=\chi^2_\mathrm{min}+n_p\ln N$, where $\chi^2_\mathrm{min}=\chi^2(\mathbf{v}_\mathrm{best})$ and $N$ is the number of SNe. Based on IC (AICc or BIC), the preferred model is the one with the smallest IC value ($\mathrm{IC}_\mathrm{min}$), and the strength of evidence for each model is given by $\Delta_\mathrm{IC}=\mathrm{IC}-\mathrm{IC}_\mathrm{min}$. Models with $\Delta_\mathrm{AICc}<2$ and $4<\Delta_\mathrm{AICc}<7$ have substantial and considerably less support, respectively \citep{Burnham_Anderson2002}, while $\Delta_\mathrm{BIC}<2$ and $6<\Delta_\mathrm{BIC}<10$ indicate little and strong evidence against the model, respectively \citep{kass1995}. Therefore, models with both $\Delta_\mathrm{AICc}<2$ and $\Delta_\mathrm{BIC}<2$ perform comparably, yielding comparable fits to the Hubble diagram, whereas models with larger $\Delta_\mathrm{AICc}$ or $\Delta_\mathrm{BIC}$ are increasingly less supported or disfavoured, respectively.

\section{Results}\label{sec:results}
\subsection{Empirical relation}\label{sec:emprical_relation}
\begin{figure}
\includegraphics[width=0.99\columnwidth]{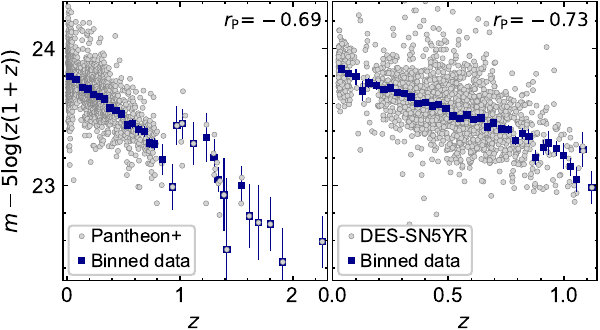}
\caption{$m-5\log(z(1+z))$ versus $z$ for Pantheon+ and DES-SN5YR. Binned data (blue squares) with error bars are shown for visualization purposes only, both here and throughout the paper.}
\label{fig:y_vs_z}
\end{figure}

Fig.~\ref{fig:y_vs_z} shows $m-5\log(z(1+z))$ versus $z$ for Pantheon+ and DES-SN5YR. Their weighted Pearson correlation coefficients ($r_\mathrm{P}$; \citealt{2019MNRAS.488.5728E}) are close to $-0.7$, with $p$-values below $10^{-220}$, indicating a moderate-to-strong negative linear correlation. Based on this, we express $m-5\log(z(1+z))$ as a linear function of $z$, with an intercept $\mathcal{M}$ and a slope $b$:
\begin{equation}\label{eq:mz_ab}
m(z)=\mathcal{M}+bz+5\log(z(1+z)).
\end{equation}
Thus, $f(z)=bz$ in Eq.~(\ref{eq:mz_ansatz}), and together with Eq.~(\ref{eq:dL_ansatz}) yields
\begin{equation}\label{eq:dL}
d_L(z)=\frac{cz}{H_0}(1+z)10^{bz/5}.
\end{equation}
We refer to Eqs.~(\ref{eq:mz_ab}) and (\ref{eq:dL}) as the empirical relation.

\begin{figure}
\includegraphics[width=0.99\columnwidth]{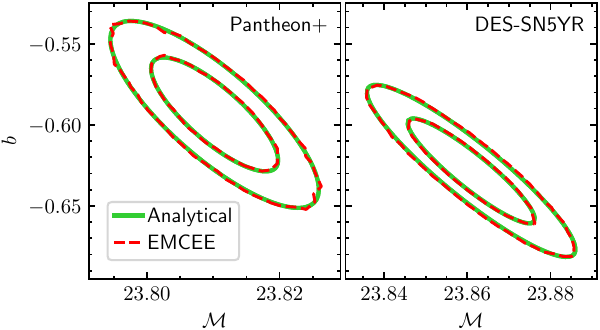}
\caption{Confidence contours at the 68.27\% and 95.45\% levels for the parameters of the empirical relation, obtained analytically and with \textsc{emcee}.} 
\label{fig:ab_pars}
\end{figure}

The empirical relation exhibits a series of characteristics. First, it is relatively stable to the absence of low-$z$ SNe (see Sect.~\ref{sec:stability}). Second, its parameters and their associated covariance matrix can be computed analytically. Fig.~\ref{fig:ab_pars} shows the confidence contours for $\mathcal{M}$ and $b$ computed analytically and with \textsc{emcee} (using the flat prior $b\in(-2,0)$), which are virtually identical. Third, under the FLRW metric, $b\propto q_0+1$. Indeed, from Eqs.~(\ref{eq:dL}) and (\ref{eq:dL_dM}),
\begin{equation}\label{eq:dM}
d_M(z)=\frac{cz}{H_0}10^{bz/5},
\end{equation}
and using Eq.~(16) of \citet{2020MNRAS.491.4960L},
\begin{equation}\label{eq:adot_a}
H(z)=H_0\frac{\sqrt{10^{-2bz/5}+\Omega_k z^2}}{1+b\ln(10) z/5}
\end{equation}
which, together with Eq.~(\ref{eq:qj}), yields
\begin{equation}\label{eq:q0_ab}
q_0=-\frac{2}{5}\ln(10)b-1.
\end{equation}

\subsection{Constraints on cosmological parameters}
Parameter constraints from Pantheon+ and DES-SN5YR for the six $m(z)$ relations used in this work are listed in Table~\ref{table:mz_pars}. The parameters for the flat $\Lambda$CDM, $\Lambda$CDM, and flat $w$CDM models are consistent with those reported by the Pantheon+ team ($\Omega_M=0.334\pm0.018$ for the flat $\Lambda$CDM model, $\Omega_M=0.306\pm0.057$ and $\Omega_\Lambda=0.625\pm0.084$ for the $\Lambda$CDM model, and $\Omega_M=0.309_{-0.069}^{+0.063}$ and $w=-0.90\pm0.14$ for the flat $w$CDM model; \citealt{2022ApJ...938..110B}) and the DES Collaboration ($\Omega_M=0.352\pm0.017$ for the flat $\Lambda$CDM model, $\Omega_M=0.291_{-0.065}^{+0.063}$ and $\Omega_\Lambda=0.55\pm0.10$ for the $\Lambda$CDM model, and $\Omega_M=0.264_{-0.096}^{+0.074}$ and $w=-0.80_{-0.16}^{+0.14}$ for the flat $w$CDM model; \citealt{2024MNRAS.533.2615C}).

\begin{table*}
  \centering
  \caption{Parameters constraints for different magnitude--redshift relations}
  \label{table:mz_pars}
  \begin{tabular}{ l c c c c c c c}
    \hline\hline
    Relation             & $\mathcal{M}$ & $b$ & $\Omega_M$ & $\Omega_\Lambda$ &$w$& $q_0$ & $\hat{j_0}$  \\
    \hline
    \multicolumn{8}{c}{DES-SN5YR} \\
    \hline
Empirical         & $23.861\pm0.010$ & $-0.628\pm0.021$ &  --                       &  --                       &  --                     &  --                        &  --                    \\
Flat $\Lambda$CDM & $23.857\pm0.011$ &  --              & $0.350\pm0.017$           &  --                       &  --                     &  --                        &  --                    \\
$\Lambda$CDM      & $23.863\pm0.013$ &  --              & $0.297_{-0.067}^{+0.060}$ & $0.566_{-0.106}^{+0.093}$ &  --                     &  --                        &  --                    \\
Flat $w$CDM       & $23.865\pm0.013$ &  --              & $0.273_{-0.101}^{+0.070}$ &  --                       & $-0.82_{-0.16}^{+0.16}$ &  --                        &  --                    \\
Pad\'e            & $23.866\pm0.014$ &  --              &  --                       &  --                       &  --                     & $-0.396_{-0.112}^{+0.072}$ & $0.664_{-0.452}^{+0.851}$ \\
Flat Pad\'e       & $23.861\pm0.011$ &  --              &  --                       &  --                       &  --                     & $-0.441\pm0.024$           &  --                    \\
    \hline
    \multicolumn{8}{c}{Pantheon+} \\
    \hline
Empirical         & $23.810\pm0.006$ & $-0.594\pm0.023$ &  --                       &  --                       &  --                     &  --                        &  --                    \\
Flat $\Lambda$CDM & $23.807\pm0.007$ &  --              & $0.331\pm0.018$           &  --                       &  --                     &  --                        &  --                    \\
$\Lambda$CDM      & $23.810\pm0.008$ &  --              & $0.298_{-0.055}^{+0.053}$ & $0.619_{-0.084}^{+0.077}$ &  --                     &  --                        &  --                    \\
Flat $w$CDM       & $23.811\pm0.009$ &  --              & $0.293_{-0.073}^{+0.067}$ &  --                       & $-0.91_{-0.17}^{+0.14}$ &  --                        &  --                    \\
Pad\'e            & $23.809\pm0.010$ &  --              &  --                       &  --                       &  --                     & $-0.492_{-0.111}^{+0.069}$ & $1.167_{-0.502}^{+0.955}$ \\
Flat Pad\'e       & $23.811\pm0.007$ &  --              &  --                       &  --                       &  --                     & $-0.471\pm0.026$           &  --                    \\
    \hline
    \multicolumn{8}{c}{Pantheon+ ($z<1.121$)} \\
    \hline
Empirical         & $23.812\pm0.007$         & $-0.606\pm0.026$         &  --                       &  --                       &  --                     &  --                        &  --                       \\
Flat $\Lambda$CDM & $23.808\pm0.007$ &  --              & $0.336\pm0.019$           &  --                       &  --                     &  --                        &  --                       \\
$\Lambda$CDM      & $23.810\pm0.009$ &  --              & $0.315_{-0.091}^{+0.082}$ & $0.636_{-0.121}^{+0.107}$ &  --                     &  --                        &  --                       \\
Flat $w$CDM       & $23.809\pm0.010$ &  --              & $0.324_{-0.117}^{+0.074}$ &  --                       & $-0.97_{-0.21}^{+0.22}$ &  --                        &  --                       \\
Pad\'e            & $23.809\pm0.010$ &  --              &  --                       &  --                       &  --                     & $-0.505_{-0.136}^{+0.082}$ & $1.323_{-0.660}^{+1.324}$ \\
Flat Pad\'e       & $23.811\pm0.007$ &  --              &  --                       &  --                       &  --                     & $-0.468\pm0.028$           &  --                       \\
    \hline
  \end{tabular}
\end{table*}

For each $m(z)$ relation, the parameters derived from DES-SN5YR and Pantheon+ are consistent within $1\,\sigma$, except for $\mathcal{M}$, which shows differences greater than $3.3\,\sigma$. The discrepancy in $\mathcal{M}$ is due to the constant offset of 0.04\,mag between selection bias corrections in DES-SN5YR and Pantheon+ \citep{2025MNRAS.541.2585V}. Accounting for this offset, the $\mathcal{M}$ values derived from DES-SN5YR and Pantheon+ become consistent within $1\,\sigma$.

\subsection{Hubble diagrams and model comparison}\label{sec:HD}
\begin{figure*}
\includegraphics[width=0.99\columnwidth]{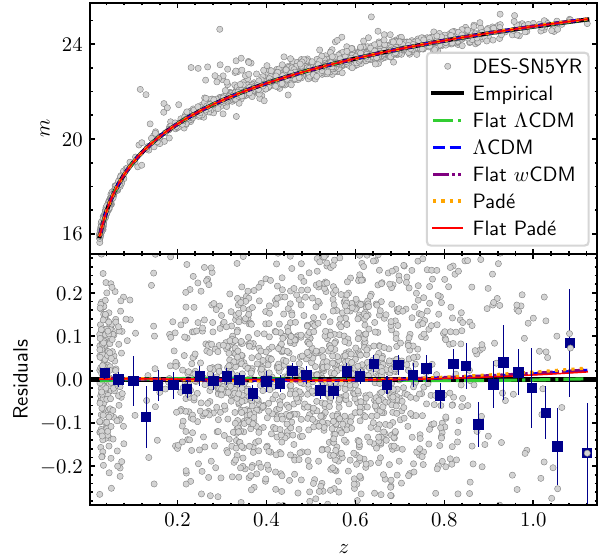}
\includegraphics[width=0.99\columnwidth]{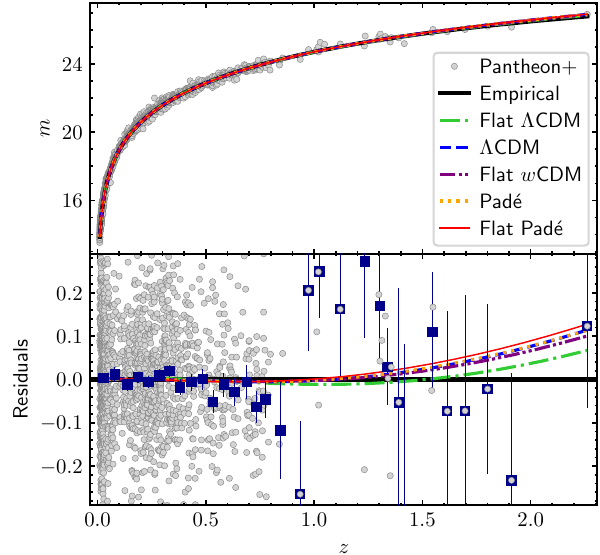}
\caption{DES-SN5YR and Pantheon+ Hubble diagrams, along with the best fits for the $m(z)$ relations. The lower panels show the residuals relative to the empirical relation.}
\label{fig:m_vs_z}
\end{figure*}

Fig.~\ref{fig:m_vs_z} shows the Hubble diagrams for DES-SN5YR and Pantheon+, the best fits for the $m(z)$ relations, and the residuals relative to the empirical relation. For DES-SN5YR, the empirical relation provides a fit comparable to that of the other $m(z)$ relations. The differences between the $m(z)$ relations and the empirical relation have rms values below 0.009\,mag and maximum absolute deviations $\Delta m_\mathrm{max}$ below 0.027\,mag. For Pantheon+, the empirical relation starts to diverge from the other $m(z)$ relations at $z\gtrsim1.4$, and for $z<1.121$ (the highest $z$ in DES-SN5YR), the rms and $\Delta m_\mathrm{max}$ remain below 0.007 and 0.012\,mag, respectively.

Given that the empirical relation accurately reproduces the Hubble diagrams of both samples up to $z=1.121$, we assume it is valid at least up to that redshift. Beyond it, its validity cannot be reliably assessed because Pantheon+ contains only 19~SNe at $z>1.121$. In what follows, we restrict our analyses of Pantheon+, including its deep-field regions, to $z<1.121$, although analyses without this cut lead to the same conclusions. Table~\ref{table:mz_pars} lists the parameter constraints for Pantheon+ ($z<1.121$).

Table~\ref{table:IC} lists the $\Delta_\mathrm{AICc}$ and $\Delta_\mathrm{BIC}$ values. Based on these, the empirical relation, the flat $\Lambda$CDM model, and the flat Pad\'e model provide comparable fits to the Hubble diagrams of DES-SN5YR and Pantheon+ ($z<1.121$), with the flat $\Lambda$CDM model being only marginally disfavoured for DES-SN5YR ($\Delta_\mathrm{AICc}=2.2$ and $\Delta_\mathrm{BIC}=2.1$). In contrast, the $\Lambda$CDM model, the flat $w$CDM model, and the Pad\'e cosmography are less supported by the AICc and disfavoured by the BIC, reflecting the penalty associated with their additional free parameter.

\begin{table}
  \centering
  \caption{AICc and BIC statistics}
  \label{table:IC}
  \begin{tabular}{l c c c c c}
    \hline\hline
   Relation          & $\chi_\mathrm{min}^2$  & AICc & BIC & $\Delta_\mathrm{AICc}$ & $\Delta_\mathrm{BIC}$\\  
    \hline
    \multicolumn{6}{c}{DES-SN5YR} \\
    \hline
Empirical         & 1638.1 & 1642.1 & 1653.2 &  0.0 &  0.0 \\
Flat $\Lambda$CDM & 1640.3 & 1644.3 & 1655.3 &  2.2 &  2.1 \\
$\Lambda$CDM      & 1639.5 & 1645.5 & 1662.1 &  3.4 &  8.9 \\
Flat $w$CDM       & 1639.0 & 1645.0 & 1661.5 &  2.9 &  8.3 \\
Pad\'e            & 1639.4 & 1645.4 & 1662.0 &  3.3 &  8.8 \\
Flat Pad\'e       & 1639.7 & 1643.7 & 1654.7 &  1.6 &  1.5 \\
    \hline
    \multicolumn{6}{c}{Pantheon+ ($z<1.121$)} \\
    \hline
Empirical         & 1392.3 & 1396.3 & 1407.0 &  0.6 &  0.5 \\
Flat $\Lambda$CDM & 1391.7 & 1395.7 & 1406.5 &  0.0 &  0.0 \\
$\Lambda$CDM      & 1391.7 & 1397.7 & 1413.7 &  2.0 &  7.2 \\
Flat $w$CDM       & 1391.7 & 1397.7 & 1413.8 &  2.0 &  7.3 \\
Pad\'e            & 1391.7 & 1397.7 & 1413.7 &  2.0 &  7.2 \\
Flat Pad\'e       & 1391.8 & 1395.8 & 1406.5 &  0.1 &  0.0 \\
    \hline 
  \end{tabular}
\end{table}

\subsection{Deceleration parameter and Hubble constant}
Table~\ref{table:q0} lists the $q_0$ values. For the flat $\Lambda$CDM, $\Lambda$CDM, and flat $w$CDM models we use $q_0=\Omega_M/2+(1+3w)\Omega_\Lambda/2$. Although the empirical relation is developed only to fit Hubble diagrams, its $q_0$ values are consistent with those of the other $m(z)$ relations.

\begin{table}
  \centering
  \caption{Values of $q_0$  for different magnitude--redshift relations}
  \label{table:q0}
  \begin{tabular}{ l c c}
    \hline\hline
    Relation       & DES-SN5YR                  & Pantheon+ $(z<1.121)$ \\
                   & $q_0$                      & $q_0$                 \\
    \hline
 Empirical         & $-0.422_{-0.019}^{+0.019}$ & $-0.442_{-0.024}^{+0.024}$ \\
 Flat $\Lambda$CDM & $-0.475_{-0.025}^{+0.026}$ & $-0.496_{-0.028}^{+0.030}$ \\
 $\Lambda$CDM      & $-0.417_{-0.067}^{+0.077}$ & $-0.478_{-0.071}^{+0.081}$ \\
 Flat $w$CDM       & $-0.392_{-0.074}^{+0.076}$ & $-0.483_{-0.092}^{+0.096}$ \\
 Pad\'e            & $-0.396_{-0.112}^{+0.072}$ & $-0.505_{-0.136}^{+0.082}$ \\
 Flat Pad\'e       & $-0.441_{-0.023}^{+0.025}$ & $-0.468_{-0.027}^{+0.028}$ \\
 \hline
   \end{tabular}
\end{table}

We estimate $H_0$ using the empirical relation together with Pantheon+ ($z<1.121$) and the Cepheid distance moduli ($\mu_\mathrm{Ceph}$) from SH0ES \citep{2022ApJ...934L...7R}. To do this, we replace Eq.~(\ref{eq:delta_m}) with $\Delta m_i=m_i-M-\mu_{\mathrm{Ceph},i}$ for those SNe hosted in galaxies with 
$\mu_\mathrm{Ceph}$. We obtain $M=-19.244\pm0.030$,  $b=-0.604\pm0.026$, and $H_0=73.4\pm1.0$\,km\,s$^{-1}$\,Mpc$^{-1}$. This $H_0$ value is consistent with those obtained by \citet{2022ApJ...938..110B} from Pantheon+ and SH0ES, which range from $73.3\pm1.1$ to $73.6\pm1.1$\,km\,s$^{-1}$\,Mpc$^{-1}$.

\subsection{Stability in the absence of low-$z$ SNe}\label{sec:stability}
\begin{figure}
\includegraphics[width=0.99\columnwidth]{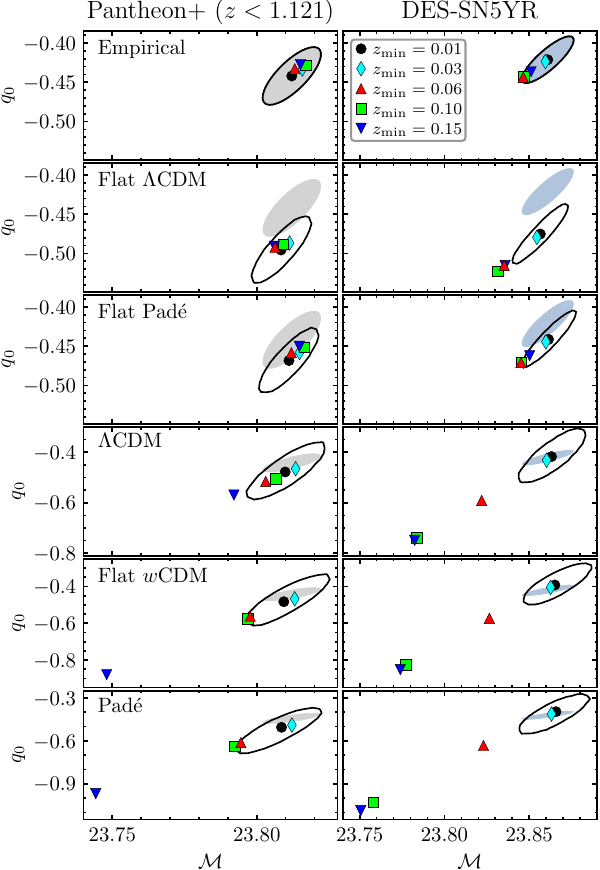}
\caption{$q_0$ versus $\mathcal{M}$ for the $m(z)$ relations fitted to DES-SN5YR and Pantheon+ ($z<1.121$) with different $z_\mathrm{min}$. Solid curves are the 68.27\% confidence contours for $z_\mathrm{min}=0.01$, while the contour of the empirical relation is shown as a shaded region in each panel for comparison. Note that, although the flat $w$CDM constraints in the ($w$, $\Omega_M$) plane are highly non-Gaussian, their projection onto $q_0$ leads to nearly symmetric uncertainties, as expected since $q_0=Q_H(z=0)$, where the parameter $Q_H(z)$ is almost free of the $w$-$\Omega_M$ degeneracy \citep{2024MNRAS.533.2615C}.}
\label{fig:a_vs_x_zmin}
\end{figure}

To evaluate the stability of the best-fit parameters of the six $m(z)$ relations against the absence of low-$z$ SNe, we compute $\mathcal{M}$ and $q_0$ for different minimum redshifts ($z_\mathrm{min}$): 0.01, 0.03, 0.06, 0.10, and 0.15. The resulting ($\mathcal{M}$, $q_0$) points are shown in Fig.~\ref{fig:a_vs_x_zmin}. We find that the empirical relation, the flat $\Lambda$CDM model, and the flat Pad\'e model are the most stable in the absence of low-$z$ SNe. Since the best-fit parameters of the empirical relation are the intercept and slope of the approximately linear correlation between $m-5\log(z(1+z))$ and $z$ (see Fig.~\ref{fig:y_vs_z} and Eq.~\ref{eq:mz_ab}), their stability depends primarily on the $z$ baseline of the data. Therefore, they are expected to vary only mildly when low-$z$ SNe are removed.

When fitted to DES-SN5YR, these three relations exhibit a gap between the cases with $z_\mathrm{min}\geq0.06$ and those with $z_\mathrm{min}\leq0.03$. Since for $z_\mathrm{min}\geq0.06$ at least 85\% of the SNe in the low-$z$ sample of DES-SN5YR are removed, one possible explanation is a difference between the $m$ values of the low-$z$ sample relative to the rest of DES-SN5YR \citep{2025SCPMA..6800413H}. This effect does not impact the analysis of the S1, S2, S4, and S5 deep-field regions, as they do not contain any SNe from the low-$z$ sample.

\subsection{Deep-field regions}
\begin{figure*}
\includegraphics[width=0.33\textwidth]{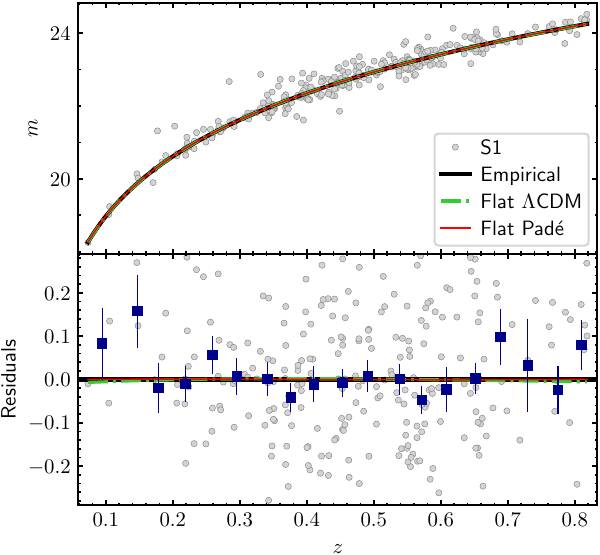}
\includegraphics[width=0.33\textwidth]{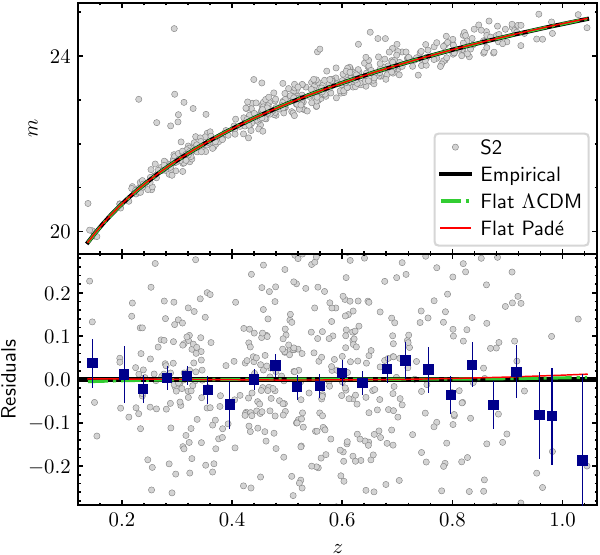}
\includegraphics[width=0.33\textwidth]{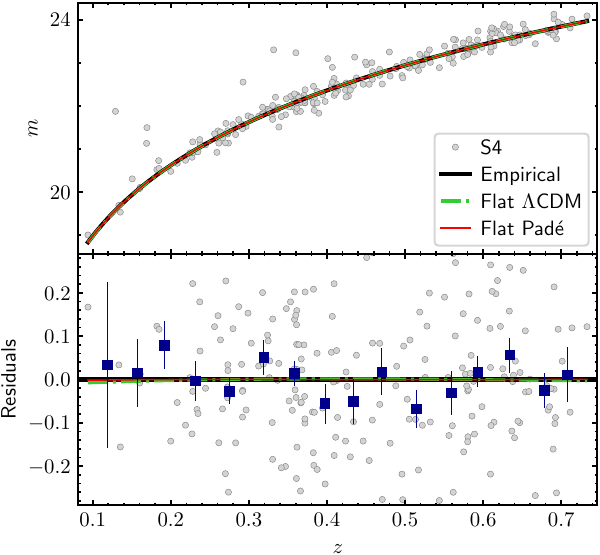}
\includegraphics[width=0.33\textwidth]{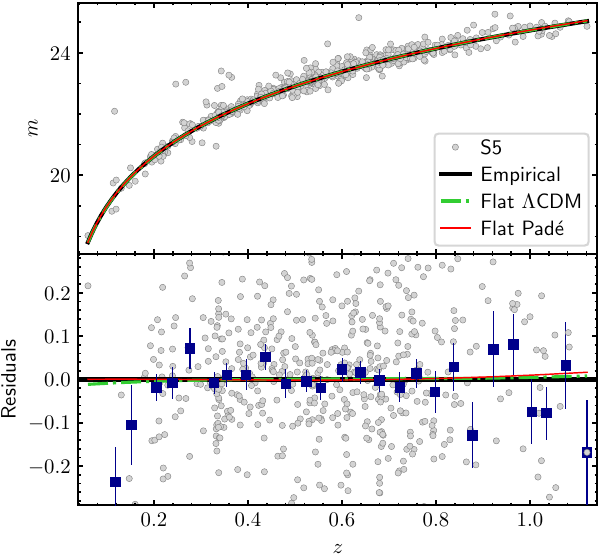}
\includegraphics[width=0.33\textwidth]{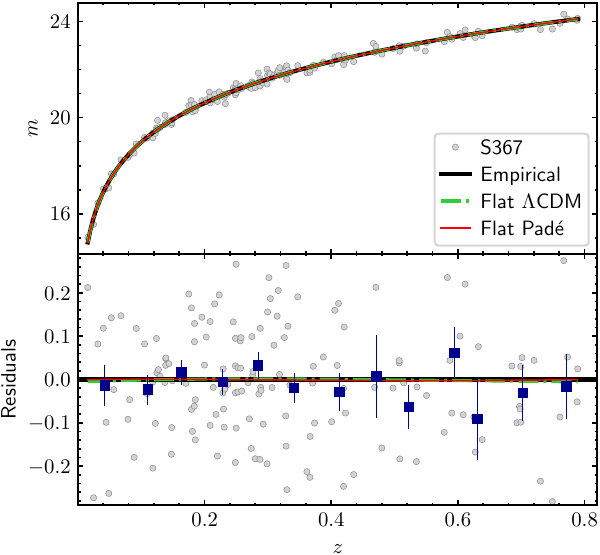}
\includegraphics[width=0.33\textwidth]{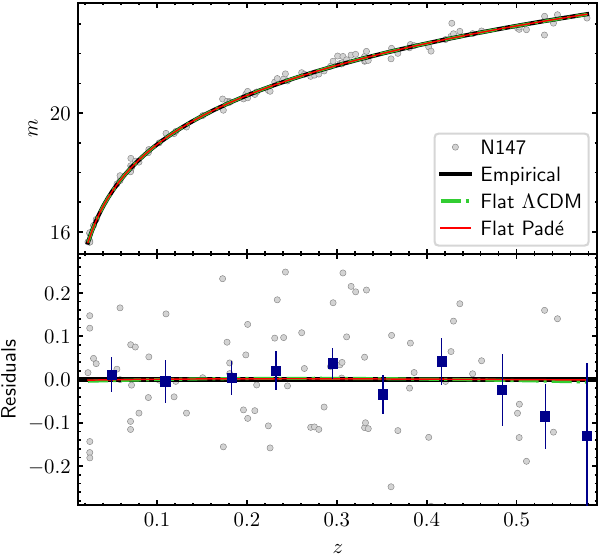}
\includegraphics[width=0.33\textwidth]{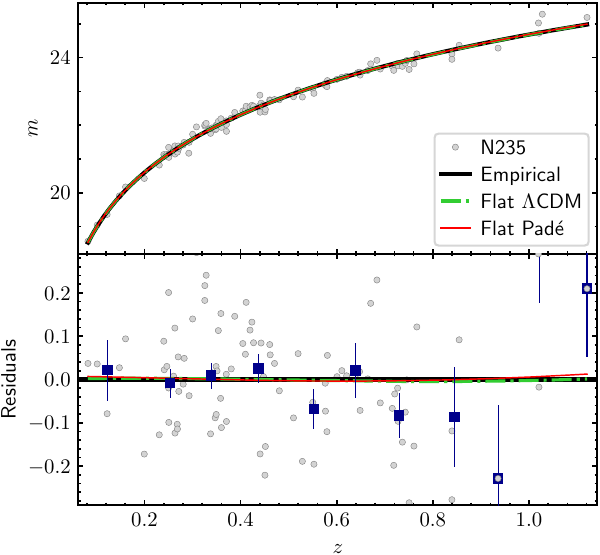}
\includegraphics[width=0.33\textwidth]{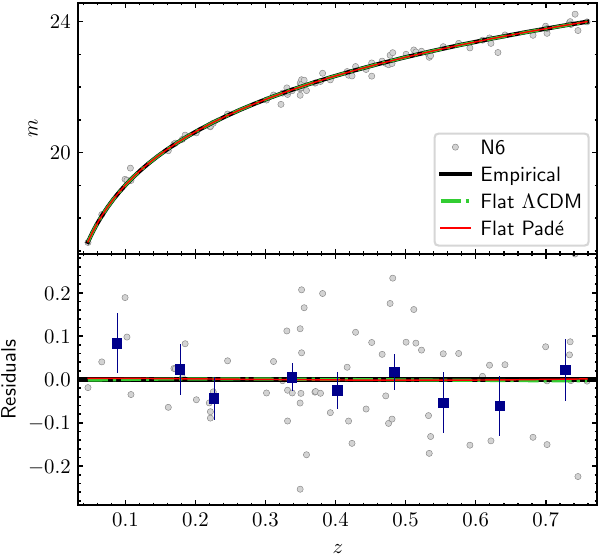}
\caption{Same as Fig.~\ref{fig:m_vs_z}, but for the deep-field regions.}
\label{fig:m_vs_z_regions} 
\end{figure*}

Fig.~\ref{fig:m_vs_z_regions} shows the Hubble diagrams for the deep-field regions. To constrain the parameters of the S1, S2, S4 and S5 regions, we select the $m$ and $z$ values for the SNe in these regions, along with the corresponding entries of the covariance matrix. Then, we replace Eq.~(\ref{eq:delta_m}) with $\Delta m_i=m_i-m(z_i,\mathbf{v}_j)$ for those SNe within the $j$-th region with parameters $\mathbf{v}_j$. The same procedure is applied to the S367, N147, N235, and N6 regions. Table~\ref{table:PD_parameters} lists the parameter constraints for the empirical relation and the flat $\Lambda$CDM and flat Pad\'e models. The best fits are shown in Fig.~\ref{fig:m_vs_z_regions}.

\begin{table*}
  \centering
  \caption{Parameters constraints for the deep-field regions}
  \label{table:PD_parameters}
  \begin{tabular}{c c c c c c c c c}
    \hline
    \hline
  Region      & \multicolumn{2}{c}{Empirical relation} & & \multicolumn{2}{c}{Flat $\Lambda$CDM model}& & \multicolumn{2}{c}{Flat Pad\'e model} \\
  \cline{2-3} \cline{5-6} \cline{8-9}
          & $\mathcal{M}$              & $b$ & &  $\mathcal{M}$ & $\Omega_M$ & & $\mathcal{M}$ & $q_0$\\
\hline
S1   & $23.823\pm0.029$ & $-0.530\pm0.059$ & & $23.810_{-0.028}^{+0.036}$ & $ 0.273_{-0.037}^{+0.049}$ & & $23.820_{-0.028}^{+0.034}$ & $-0.548_{-0.056}^{+0.070}$ \\
S2   & $23.850\pm0.022$ & $-0.622\pm0.043$ & & $23.837_{-0.024}^{+0.029}$ & $ 0.333_{-0.031}^{+0.042}$ & & $23.851_{-0.023}^{+0.027}$ & $-0.448_{-0.047}^{+0.057}$ \\ 
S4   & $23.821\pm0.032$ & $-0.511\pm0.069$ & & $23.806_{-0.030}^{+0.038}$ & $ 0.257_{-0.040}^{+0.055}$ & & $23.816_{-0.031}^{+0.037}$ & $-0.571_{-0.064}^{+0.079}$ \\
S5   & $23.847\pm0.021$ & $-0.618\pm0.038$ & & $23.831_{-0.022}^{+0.027}$ & $ 0.327_{-0.028}^{+0.036}$ & & $23.846_{-0.021}^{+0.025}$ & $-0.455_{-0.041}^{+0.049}$ \\
\hline
S367 & $23.834\pm0.023$ & $-0.609\pm0.069$ & & $23.828_{-0.022}^{+0.027}$ & $ 0.335_{-0.045}^{+0.059}$ & & $23.833_{-0.022}^{+0.026}$ & $-0.465_{-0.067}^{+0.083}$ \\
N147 & $23.781\pm0.031$ & $-0.436\pm0.111$ & & $23.775_{-0.028}^{+0.035}$ & $ 0.227_{-0.060}^{+0.078}$ & & $23.778_{-0.029}^{+0.034}$ & $-0.637_{-0.092}^{+0.114}$ \\
N235 & $23.830\pm0.036$ & $-0.654\pm0.076$ & & $23.830_{-0.038}^{+0.050}$ & $ 0.376_{-0.056}^{+0.084}$ & & $23.838_{-0.037}^{+0.046}$ & $-0.397_{-0.083}^{+0.111}$ \\
N6   & $23.865\pm0.042$ & $-0.652\pm0.103$ & & $23.862_{-0.040}^{+0.055}$ & $ 0.370_{-0.068}^{+0.103}$ & & $23.868_{-0.042}^{+0.051}$ & $-0.412_{-0.103}^{+0.140}$ \\
    \hline 
  \end{tabular}
\end{table*}

Table~\ref{table:IC_regions} lists the $\Delta_\mathrm{AICc}$ and $\Delta_\mathrm{BIC}$ values for the empirical relation and the flat $\Lambda$CDM and flat Pad\'e models. These values indicate that the three $m(z)$ relations provide comparable fits to the Hubble diagrams in the deep-field regions. Since the empirical relation is purely phenomenological and its parameters, uncertainties, and confidence contours can be computed analytically (Sect.~\ref{sec:emprical_relation}), we use it as a convenient parametrization to characterize the deep-field regions.

\begin{table}
  \centering
  \caption{AICc and BIC statistics for the deep-field regions}
  \label{table:IC_regions}
  \begin{tabular}{l c c c c c}
    \hline\hline
   Relation            & $\chi^2_\mathrm{min}$ & AICc & BIC & $\Delta_\mathrm{AICc}$ & $\Delta_\mathrm{BIC}$\\
    \hline
  \multicolumn{6}{c}{S367+N147+N235+N6} \\
  \hline
Empirical         & 376.8 & 393.1 & 425.3 & 0.6 & 0.7\\
Flat $\Lambda$CDM & 376.2 & 392.5 & 424.7 & 0.0 & 0.1\\
Flat Pad\'e       & 376.1 & 392.5 & 424.6 & 0.0 & 0.0\\
    \hline    
  \multicolumn{6}{c}{S1+S2+S4+S5} \\
  \hline
Empirical         & 1442.9 & 1459.0 & 1502.1 & 0.0 & 0.0 \\
Flat $\Lambda$CDM & 1443.9 & 1460.0 & 1503.1 & 1.0 & 1.0 \\
Flat Pad\'e       & 1444.6 & 1460.7 & 1503.8 & 1.7 & 1.7 \\
    \hline
  \end{tabular}
\end{table}

\begin{figure}
\includegraphics[width=0.99\columnwidth]{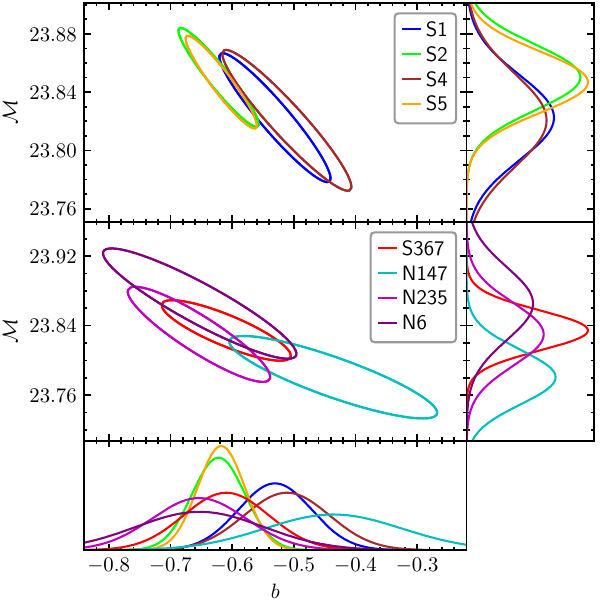}
\caption{Confidence contours (68.27\%) and marginalized distributions for the parameters of the empirical relation fitted to the deep-field regions.}
\label{fig:a_vs_b_regions}
\end{figure}

Fig.~\ref{fig:a_vs_b_regions} shows the confidence contours and the marginalized distributions for $\mathcal{M}$ and $b$ for the deep-field regions. The $b$ values are consistent within 1.6\,$\sigma$. When accounting for the 0.04\,mag offset between the selection bias corrections in DES-SN5YR and Pantheon+, the $\mathcal{M}$ values are consistent within 1.6\,$\sigma$, in agreement with isotropy. Similar results are obtained when comparing the parameters of the flat $\Lambda$CDM and flat Pad\'e models, where the differences in $\mathcal{M}$, $\Omega_M$, and $q_0$ are no more than $1.7\,\sigma$, $1.5\,\sigma$ and $1.7\,\sigma$, respectively.

To test whether fitting the S1, S2, S4, and S5 regions with different parameters for each region (eight parameters in total) provides a significantly better fit than an isotropic fit (two parameters for all regions), we perform a likelihood ratio test (LRT). The test statistic $\lambda = \chi^2_\mathrm{min,iso}-\chi^2_\mathrm{min}$, where $\chi^2_\mathrm{min}$ is the value for the model with eight parameters (listed in Table~\ref{table:IC_regions}) and $\chi^2_\mathrm{min,iso}$ is the value for the isotropic fit, follows a $\chi^2$ distribution with six degrees of freedom \citep{Wilks1938}. Fitting the empirical relation, we obtain $\chi^2_\mathrm{min,iso}=1451.3$, $\lambda=8.4$, and a $p$-value of 0.21. For the S367, N235, N147, and N6 regions, we obtain $\chi^2_\mathrm{min,iso}=383.5$, $\lambda=6.7$, and a $p$-value of 0.35. In both cases, the isotropic fit cannot be rejected. Similar results are obtained when using the flat $\Lambda$CDM and flat Pad\'e models, where the LRT $p$-values are between 0.20 and 0.36. Therefore, we find no evidence for anisotropy in the Hubble diagrams of the deep-field regions.

Table~\ref{table:q0j0_regions} lists the $q_0$ values for the deep-field regions. The values measured using the empirical relation are consistent within 1.6\,$\sigma$ and lower than zero by more than 4.8\,$\sigma$ (99.9998\% confidence). These results indicate that the Universe is undergoing accelerated expansion in a statistically consistent manner across all eight deep-field regions, under the assumption that the large-scale densities in these regions are independent of the comoving radial coordinate.

\begin{table}
  \centering
  \caption{Values of $q_0$ for the deep-field regions.}
  \label{table:q0j0_regions}
  \begin{tabular}{c c c c}
    \hline\hline
 Region & $q_0$(Empirical) & $q_0$(Flat Pad\'e) & $q_0$(Flat $\Lambda$CDM)\\
 \hline
 S1   & $-0.511\pm0.055$ & $-0.548_{-0.056}^{+0.070}$ & $-0.591_{-0.055}^{+0.075}$ \\
 S2   & $-0.427\pm0.040$ & $-0.448_{-0.047}^{+0.057}$ & $-0.500_{-0.047}^{+0.062}$ \\
 S4   & $-0.529\pm0.064$ & $-0.571_{-0.064}^{+0.079}$ & $-0.615_{-0.060}^{+0.083}$ \\
 S5   & $-0.431\pm0.035$ & $-0.455_{-0.041}^{+0.049}$ & $-0.510_{-0.041}^{+0.055}$ \\
 \hline
 S367 & $-0.439\pm0.064$ & $-0.465_{-0.067}^{+0.083}$ & $-0.497_{-0.067}^{+0.089}$ \\
 N147 & $-0.598\pm0.102$ & $-0.637_{-0.092}^{+0.114}$ & $-0.660_{-0.089}^{+0.117}$ \\
 N235 & $-0.398\pm0.070$ & $-0.397_{-0.083}^{+0.111}$ & $-0.436_{-0.084}^{+0.126}$ \\
 N6   & $-0.399\pm0.095$ & $-0.412_{-0.103}^{+0.140}$ & $-0.445_{-0.103}^{+0.154}$ \\
 \hline
  \end{tabular}
\end{table}

\subsection{Application to other samples}

\subsubsection{DES-Dovekie sample}
While this work was nearing completion, a re-analysis of DES-SN5YR was presented by \citet{2025arXiv251107517P}. The resulting recalibrated data set, DES-Dovekie,\footnote{\url{https://github.com/des-science/DES-SN5YR}} includes an improved photometric cross-calibration and a fixed host-galaxy color law. Since DES-Dovekie supersedes DES-SN5YR \citep{2025arXiv251107517P}, its analysis provides the best constraints and overall results from DES-SN combined with a low-$z$ sample.

We fit the empirical relation ($b=-0.603\pm0.020$), the Pad\'e cosmography ($q_0=-0.444_{-0.114}^{+0.072}$, $\hat{j_0}=0.805_{-0.466}^{+0.881}$), flat $\Lambda$CDM ($\Omega_M=0.329\pm0.015$), $\Lambda$CDM ($\Omega_M=0.286_{-0.062}^{+0.056}$, $\Omega_\Lambda=0.600_{-0.102}^{+0.088}$ or $\Omega_k=0.11_{-0.14}^{+0.16}$), flat $w$CDM ($\Omega_M=0.268_{-0.086}^{+0.063}$, $w=-0.85\pm0.15$), and the flat Pad\'e model ($q_0=-0.470\pm0.022$). The parameters are consistent with those from DES-SN5YR and from Pantheon+ ($z<1.121$), as well as with those reported by \citet{2025arXiv251107517P} for the flat $\Lambda$CDM ($\Omega_M=0.330\pm0.015$), $\Lambda$CDM ($\Omega_M=0.279\pm0.057$, $\Omega_k=0.14\pm0.15$), and flat $w$CDM model ($\Omega_M=0.263_{-0.078}^{+0.064}$, $w=-0.838_{-0.142}^{+0.130}$). The $\Delta_\mathrm{AICc}$ and $\Delta_\mathrm{BIC}$ values are similar to those for DES-SN5YR.

It is worth noting that the magnitude differences between a flat $\Lambda$CDM model with $\Omega_M$ from Planck CMB measurements ($\Omega_M^\mathrm{Planck}=0.3166$; \citealt{2020AA...641A...6P}) and the empirical relation have $\mathrm{rms}=0.004$\,mag and a maximum absolute deviation of $\Delta m_\mathrm{max}=0.011$\,mag, while the $(\mathrm{rms},\,\Delta m_\mathrm{max})$ values for the empirical relation calibrated with DES-SN5YR and Pantheon+ ($z<1.121$) are $(0.012,\,0.025)$ and $(0.008,\,0.014)$, respectively. Therefore, the empirical relation shows a close agreement with the flat $\Lambda$CDM model using $\Omega_M^\mathrm{Planck}$, particularly when calibrated with DES-Dovekie and Pantheon+ ($z<1.121$).

The $b$ values for the S1, S2, S4, and S5 regions in DES-Dovekie are $-0.555\pm0.058$, $-0.612\pm0.040$, $-0.474\pm0.066$, and $-0.585\pm0.036$, respectively. These values, together with those from S367, N147, N235, and N6, are mutually consistent within $1.8\,\sigma$. Using a LRT, we obtain $\lambda=6.6$ and a $p$-value of 0.36, indicating that the isotropic fit cannot be rejected. The $b$ values from the S1, S2, S4, and S5 regions correspond to $q_0$ values of $-0.489\pm0.054$, $-0.436\pm0.037$, $-0.564\pm0.061$, and $-0.461\pm0.033$, respectively.

We also constrain parameters using only DES SNe within DES-Dovekie. To do so, we select SNe with $z>0.093$, which excludes the entire low-$z$ sample. We obtain $b=-0.589\pm0.026$ for the empirical relation, $\Omega_M=0.307\pm0.020$ for the flat $\Lambda$CDM model, and $q_0=-0.488\pm0.029$ for the flat Pad\'e model. The corresponding $\Delta$IC values are 0.5, 0.0, and 2.1, indicating that the empirical relation and the flat $\Lambda$CDM model provide comparable fits to the Hubble diagram.

\subsubsection{Amalgame sample}
\begin{figure}
\includegraphics[width=0.99\columnwidth]{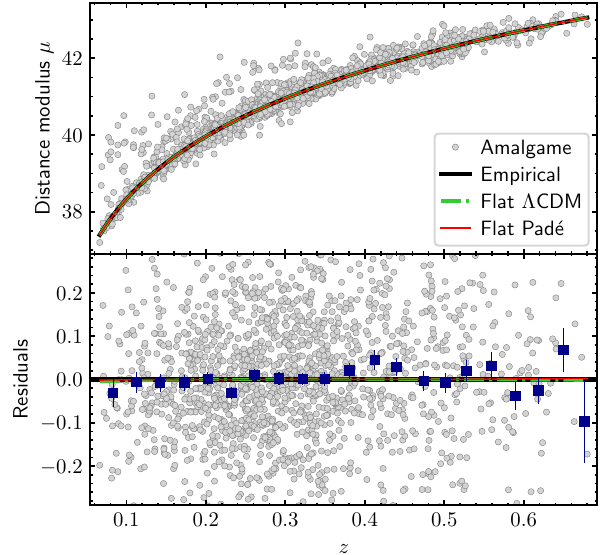}
\caption{Same as Fig.~\ref{fig:m_vs_z}, but for the Amalgame SN sample.}
\label{fig:m_vs_z_Amalgame} 
\end{figure}

The Amalgame sample \citep{2024MNRAS.529.2100P}\footnote{\url{https://github.com/bap37/AmalgameDR}} consists of 1792 photometrically classified SNe Ia from SDSS \citep{2011ApJ...738..162S} and Pan-STARRS1 \citep{2018ApJ...859..101S}, with $0.066<z<0.680$. We obtain $b=-0.613\pm0.041$ for the empirical relation, $\Omega_m=0.334_{-0.027}^{+0.031}$ for the flat $\Lambda$CDM model, and $q_0=-0.469_{-0.041}^{+0.045}$ for the flat Pad\'e model. These values are consistent with those from Pantheon+ ($z<1.121$), DES-SN5YR, and DES-Dovekie. Fig.~\ref{fig:m_vs_z_Amalgame} shows the Hubble diagram for Amalgame. The $\Delta$IC values for the empirical relation, the flat $\Lambda$CDM model, and the flat Pad\'e model are 0.4, 0.0, and 0.6, respectively, indicating that they provide comparable fits to the Hubble diagram.

\subsubsection{BAO distances and cosmic chronometers}
Although the empirical relation is designed to fit Hubble diagrams of standard candles, it is worth applying it to available $d_M(z)$ and $H(z)$ datasets. To do so, we assume the FLRW metric to transform the empirical relation into expressions for $d_M(z)$ (Eq.~\ref{eq:dM}) and $H(z)$ (Eq.~\ref{eq:adot_a}).

\citet{2025PhRvD.112h3515A} reported 13 distances measured from baryon acoustic oscillations (BAO) in DESI DR2, relative to the sound horizon $r_d$, at effective redshifts $z_\mathrm{eff}$ ranging from 0.295 to 2.33: six $d_M(z)/r_d$ values, six $d_H(z)/r_d$ values (with $d_H(z)=c/H(z)$), and one $d_V(z)/r_d$ value, where $d_V(z)\equiv[z d_M(z)^2 d_H(z)]^{1/3}$.\footnote{Data are available at \url{https://github.com/CobayaSampler/bao_data/tree/master/desi_bao_dr2}} Fitting Eqs.~(\ref{eq:dM}) and (\ref{eq:adot_a}) to the seven BAO distances with $z_\mathrm{eff}<1.121$, and assuming $\Omega_k=0$, we obtain $b=-0.584\pm0.016$. For comparison, the $b$ values obtained from Pantheon+ ($z<1.121$), DES-SN5YR, DES-Dovekie, and Amalgame, when all SN samples are restricted to $z\geq0.295$, are $-0.697\pm0.050$, $-0.622\pm0.035$, $-0.618\pm0.033$, and $-0.688\pm0.070$, respectively. Compared to the BAO constraint, the latter three values are consistent within $1.4\,\sigma$, while the Pantheon+ ($z<1.121$) result is marginally consistent, showing a $2.2\,\sigma$ difference.

\citet{2026JHEAp..4900444A} collected 34 $H(z)$ values measured with cosmic chronometers, with $0.07\leq z\leq 1.965$. Fitting Eq.~(\ref{eq:adot_a}) to the 27 $H(z)$ values with $z<1.121$, and assuming $\Omega_k=0$, we obtain $b=-0.635\pm0.095$. This value is consistent with those from Pantheon+ ($z<1.121$), DES-SN5YR, DES-Dovekie, and Amalgame when all SN samples are restricted to $z\geq0.07$, yielding $b=-0.617\pm0.033$, $-0.607\pm0.027$, $-0.591\pm0.024$, and $-0.613\pm0.041$, respectively.

\section{Discussion}\label{sec:discussion}

\subsection{Jerk parameter}\label{sec:j0}
The value of $j_0$ provides an observational test for the flat $\Lambda$CDM model \citep{2015ApJ...814....7B}. Indeed, for the non-flat $w$CDM model (Eq.~\ref{eq:Hz}), Eq.~(\ref{eq:qj}) yields $j_0=1-\Omega_k+3(1+w)(2q_0+\Omega_k-1)/2$, which reduces to $j_0=1$ for the flat $\Lambda$CDM model. Therefore, if $j_0$ deviates from unity, it could indicate that the Universe is not flat, dark energy is not a cosmological constant, or that Eq.~(\ref{eq:Hz}) is incorrect. Given that $\hat{j_0} = 1$ for the flat $\Lambda$CDM model, the test remains valid if $\hat{j_0}$ is used instead of $j_0$.

\begin{table}
  \centering
  \caption{Values of $\hat{j_0}$ for different magnitude--redshift relations}
  \label{table:j0}
  \begin{tabular}{ l c c}
    \hline\hline
    Relation   & Pantheon+ ($z<1.121$)    & DES-SN5YR \\
               & $\hat{j_0}$     & $\hat{j_0}$  \\
    \hline
 Empirical         & $ 0.585_{-0.011}^{+0.014}$ & $ 0.596_{-0.011}^{+0.012}$  \\
 $\Lambda$CDM      & $ 0.902_{-0.415}^{+0.371}$ & $ 0.726_{-0.339}^{+0.298}$  \\
 Flat $w$CDM       & $ 0.909_{-0.575}^{+0.672}$ & $ 0.514_{-0.345}^{+0.409}$  \\
 Pad\'e            & $ 1.323_{-0.660}^{+1.324}$ & $ 0.664_{-0.452}^{+0.851}$  \\
    \hline
  \end{tabular}
\end{table}

Table~\ref{table:j0} lists the $\hat{j_0}$ values. For the empirical relation we use
\begin{equation}
\hat{j_0}=\frac{9}{4} \left(\frac{5}{9}+q_0\right)^2+\frac{5}{9},
\end{equation}
which is derived from Eqs.~(\ref{eq:qj}) and (\ref{eq:adot_a}). For DES-SN5YR and Pantheon+ ($z<1.121$), the $\hat{j_0}$ values are consistent within $1.2\,\sigma$. In particular, the $\hat{j_0}$ values for the $\Lambda$CDM model, the flat $w$CDM model, and the Pad\'e cosmography are consistent with unity within $1.2\,\sigma$. In contrast, the $\hat{j_0}$ values for the empirical relation are approximately $0.4$ below unity, with an overwhelming significance of at least $30\,\sigma$. Assuming $w=-1$, these values translate to $\Omega_k=0.208_{-0.007}^{+0.005}$ and $0.202_{-0.006}^{+0.005}$ for Pantheon+ ($z<1.121$) and DES-SN5YR, respectively; or, assuming $\Omega_k=0$, to $w=-0.853\pm0.001$ and $-0.854\pm0.001$, respectively. Note, however, that the empirical relation is intended to fit the Hubble diagram, not its derivatives. Therefore, although the empirical relation provides $q_0$ values consistent with those from other $m(z)$ relations (see Table~\ref{table:q0}), it may misestimate $\hat{j_0}$.

\subsection{Cosmic anisotropies through hemisphere comparison}
Pantheon+ has been widely used to search for anisotropies in cosmological parameters using the hemisphere comparison (HC) method \citep{2007AA...474..717S} under different cosmological models \citep[e.g.,][]{2023PhRvD.108l3533M,2024ApJ...971...19C,2024AA...681A..88H,2024AA...689A.215H}. The HC method splits the sky into two opposite hemispheres (`up' and `down'), constrains the cosmological parameter $x$ in each, computes a statistic such as the signal-to-noise ratio (S/N) of $\Delta x=|x_\mathrm{up}-x_\mathrm{down}|$, and repeats the procedure for different directions across the sky.

\begin{figure}
\includegraphics[width=0.99\columnwidth]{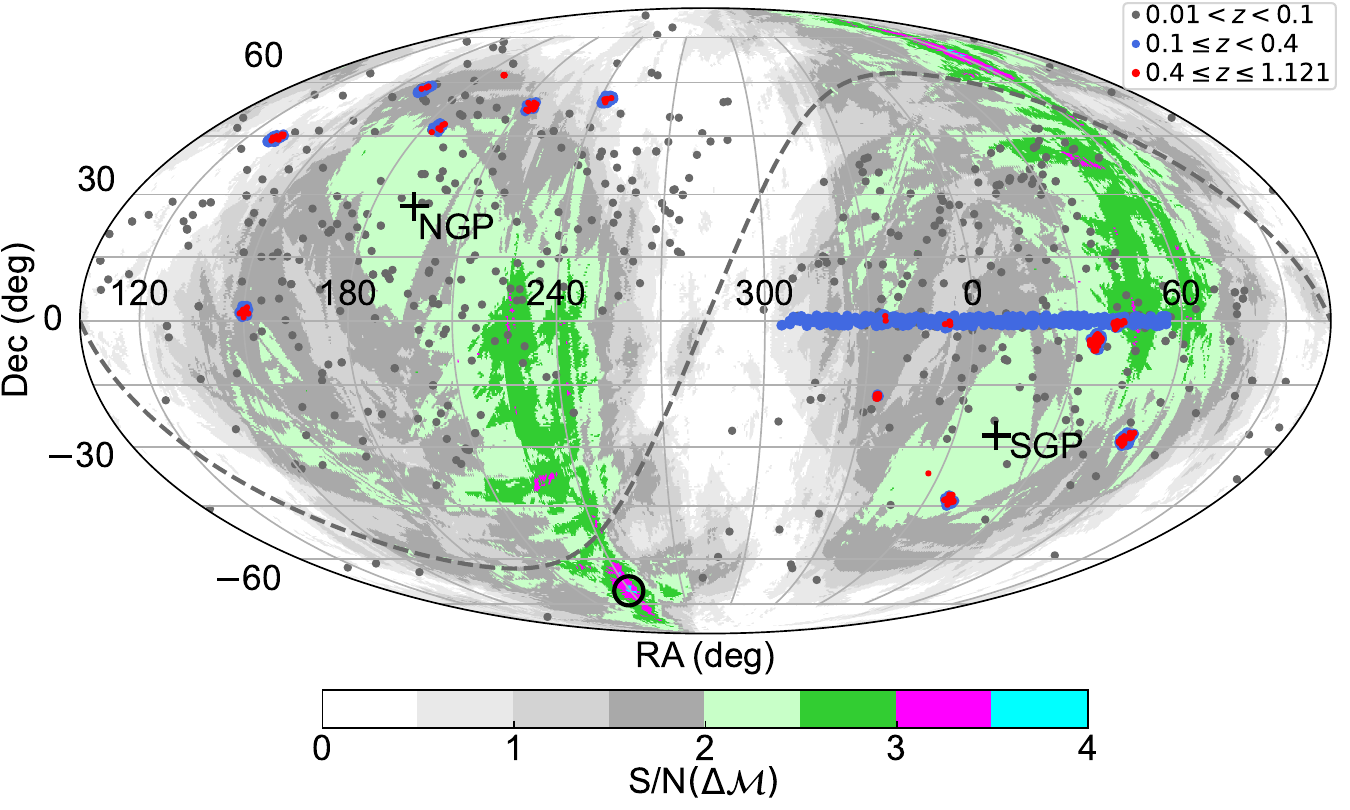}
\includegraphics[width=0.99\columnwidth]{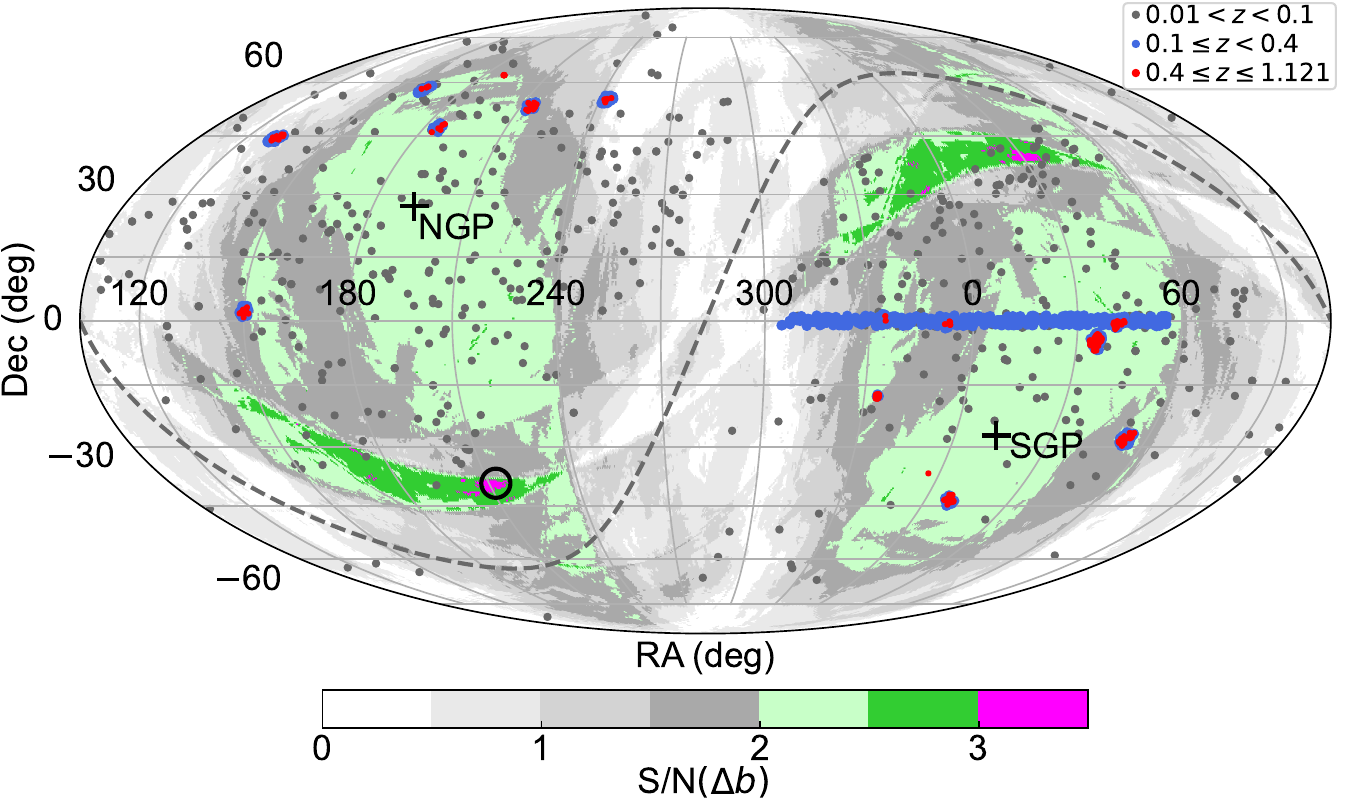}
\caption{Sky maps of $\mathrm{S/N}(\Delta \mathcal{M})$ and $\mathrm{S/N}(\Delta b)$. Dots represent the SNe in Pantheon+ ($z<1.121$) and circles mark the maximum S/N values.}
\label{fig:snr_a_b}
\end{figure}

We apply the HC method to Pantheon+ ($z<1.121$) using the empirical relation. We select a set of directions on the celestial sphere using \texttt{HEALPY} \citep{2005ApJ...622..759G,2019JOSS....4.1298Z} with a grid resolution parameter $N_\mathrm{side}=128$. This provides 196.608 equal-area pixels on the sky, corresponding to a set of 98.304 independent directions. For each direction, we compute $\mathrm{S/N}(\Delta x)=\Delta x/(\sigma_\mathrm{up}^2+\sigma_\mathrm{down}^2-2\sigma_\mathrm{up,down})^{1/2}$, where $\sigma_\mathrm{up}$ and $\sigma_\mathrm{down}$ are the uncertainties of $x_\mathrm{up}$ and $x_\mathrm{down}$, respectively, and $\sigma_\mathrm{up,down}$ is their covariance. Fig.~\ref{fig:snr_a_b} shows the sky maps of $\mathrm{S/N}(\Delta \mathcal{M})$ and $\mathrm{S/N}(\Delta b)$. Their maximum values are 3.7 at (RA, Dec)=(239.2\degr, $-70.17$\degr) and 3.3 at (RA, Dec)=(212.3\degr, $-39.07$\degr), respectively.

To evaluate the probabilities of obtaining $\mathrm{S/N}(\Delta\mathcal{M})\geq3.7$ and $\mathrm{S/N}(\Delta b)\geq3.3$ by chance when scanning many directions on the sky (the look-elsewhere effect; \citealt{2020JCAP...10..009B}), we perform $10^4$ simulations. In each simulation, we randomize the RA and Dec coordinates of the SNe and produce the $\mathrm{S/N}(\Delta \mathcal{M})$ and $\mathrm{S/N}(\Delta b)$ maps with the HC method and the same directions used above. The probabilities of obtaining $\mathrm{S/N}(\Delta \mathcal{M})\geq3.7$ and $\mathrm{S/N}(\Delta b)\geq3.3$ by chance are 6.0\% and 11.2\%, respectively, corresponding to significances of $1.9\,\sigma$ and $1.6\,\sigma$. Therefore, we find no evidence for anisotropies in $\mathcal{M}$ and $b$.

If the directions of maximum $\mathrm{S/N}(\Delta\mathcal{M})$ and $\mathrm{S/N}(\Delta b)$ were real anisotropies, detecting them at a $3\,\sigma$ level with the HC method would require adding 2500 and 4400 SNe to Pantheon+ ($z<1.121$), respectively. Another alternative would be to perform deep-field SN surveys toward the directions of the candidate anisotropies. This is particularly important given the absence of high-$z$ SNe near those directions (see Fig.~\ref{fig:snr_a_b}).

\subsection{JWST, LSST, CSST, and Roman}
Given the small number of SNe at $z>1.121$, the validity of the empirical relation beyond this redshift remains to be verified. Currently, the James Webb Space Telescope (JWST) is observing SNe~Ia at $z>1.7$ \citep{2023ApJ...954...31C,2025ApJ...979..250D}. \citet{2025arXiv251219783S} reported $\mu=46.08_{-0.18}^{+0.19}$ for SN~2025ogs at $z=2.05$, while \citet{2025ApJ...981L...9P} reported $\mu=46.10_{-0.18}^{+0.17}$ for SN~2023aeax at $z=2.15$ and $\mu=47.14_{-0.24}^{+0.21}$ for SN~2023adsy at $z=2.903$. At these redshifts, using the parameters inferred from Pantheon+ (Table~\ref{table:mz_pars}) and $H_0=70$\,km\,s$^{-1}$\,Mpc$^{-1}$, the predictions of the empirical relation are $\mu=45.92$, 46.04, and 46.71, respectively, while those of the $\Lambda$CDM model are $\mu=46.0$, 46.13, and 46.94, respectively. The $\mu$ values for SNe~2025ogs and 2023aeax are consistent with both relations, whereas that for SN~2023adsy is more consistent with the $\Lambda$CDM model than with the empirical relation. However, due to its red color, it is still unclear whether SN~2023adsy is representative of high-$z$ SNe~Ia or an outlier \citep{2024ApJ...971L..32P,2025ApJ...981L...9P}.

In addition to JWST, the deep-field surveys of the future China Space Station Telescope \citep[CSST;][]{2023SCPMA..6629511L} and the Nancy Grace Roman Space Telescope \citep{2023arXiv230702670H} will observe SNe~Ia up to $z=1.3$ and $z=1.7$, respectively. These SNe, together with those observed by JWST and the 19 SNe at $1.121<z\leq2.261$ in Pantheon+, will be crucial for determining the highest $z$ at which the empirical relation remains valid.

The deep rolling surveys of the Vera C. Rubin Observatory Legacy Survey of Space and Time \citep[LSST;][]{2024ApJS..275...21G} will provide photometry of thousands of SNe~Ia with $z\lesssim1.1$. Therefore, the Hubble diagram for each of the five LSST Deep Drilling Fields can be analyzed with the empirical relation. Of these fields, the newly defined Euclid Deep Field South (RA=61.241\degr, Dec=$-48.423$\degr), located at an angular separation of 21° from the S5 field, will allow us to study cosmic acceleration in a different direction from those analyzed in this work.

\section{Conclusions}\label{sec:conclusions}
We have presented a new empirical $m(z)$ relation based on data from Pantheon+ and DES-SN5YR. This relation fits their Hubble diagrams with only two parameters, $\mathcal{M}$ and $b$, both of which can be computed analytically. In particular, under the FLRW metric, $b\propto q_0+1$.

For the DES-SN5YR and Pantheon+ Hubble diagrams, the empirical relation provides fits up to $z=1.121$ that closely match those obtained with the Pad\'e cosmography and the $\Lambda$CDM, flat $\Lambda$CDM, flat $w$CDM, and flat Pad\'e models. The validity of the empirical relation beyond $z=1.121$ cannot be reliably assessed given the small number of SNe at $z>1.121$. Based on the $\Delta_\mathrm{IC}$ values, the empirical relation performs better than the $\Lambda$CDM model, the flat $w$CDM model, and the Pad\'e cosmography, and comparably to the flat $\Lambda$CDM and flat Pad\'e models.

For DES-SN5YR and Pantheon+ ($z<1.121$), we obtain $b=-0.628\pm0.021$ and $b=-0.606\pm0.026$, respectively. We also applied the empirical relation to DES-Dovekie and Amalgame, obtaining $b=-0.603\pm0.020$ and $b=-0.613\pm0.041$, respectively. For Pantheon+ ($z<1.121$) combined with SH0ES, we obtain $H_0=73.4\pm1.0$\,km\,s$^{-1}$\,Mpc$^{-1}$.

The empirical relation, as well as the flat $\Lambda$CDM and the flat Pad\'e models, is relatively stable in the absence of low-$z$ SNe. We fit the empirical relation to the Hubble diagrams of the eight deep-field regions, finding that their fitted $\mathcal{M}$ and $b$ parameters are consistent within $1.6\,\sigma$, 
in agreement with isotropy. Applying a LRT, we find no evidence of anisotropy. The latter is further supported by using the empirical relation in the HC method applied to Pantheon+ ($z<1.121$), which finds no statistically significant evidence for anisotropies in $\mathcal{M}$ and $b$.

Furthermore, the $b$ values from the eight deep-field regions translate to $q_0$ values ranging from $-0.6$ to $-0.4$, which are consistent within $1.6\,\sigma$ and lower than zero at $4.8\,\sigma$. These results strongly support an accelerating universe, under the assumption that the large-scale density in each deep-field region is independent of the comoving radial coordinate.

Under the assumption of the FLRW metric, the empirical relation can also be applied to BAO distances from DESI DR2, and $H(z)$ values from cosmic chronometers. The derived $b$ values, assuming $\Omega_k=0$, are mostly consistent with those from DES-SN5YR, Pantheon+ ($z<1.121$), DES-Dovekie, and Amalgame.

The empirical relation $m(z)=\mathcal{M}+bz+5\log(z(1+z))$, or equivalently $d_L(z)=\frac{cz}{H_0}(1+z)10^{bz/5}$, provides a straightforward quantitative tool for fitting Hubble diagrams of SNe~Ia without the need to add a low-$z$ sample. It enables tests of basic properties of the Universe in a simplified manner, reducing the computational cost typically associated with physically motivated cosmological models. Our analysis shows that the empirical relation is valid up to at least $z\approx1.1$. Ongoing and upcoming deep SN surveys will allow us to test whether it holds at higher redshifts.

\begin{acknowledgements}

We acknowledge the Director of the Millennium Institute of Astrophysics (MAS), Francisco F\"orster, for financial assistance through ANID grant ICN12\_009, and acknowledge complementary funding from ANID/FONDECYT grant 1251692. OR is supported by the Rubin-Chile Fund under grant DIA2650. We thank Felipe Asenjo and Ariel Órdenes for helpful discussions.

\end{acknowledgements}

\bibliographystyle{aa}
\bibliography{references}

@ARTICLE{1998AJ....116.1009R,
       author = {{Riess}, Adam G. and {Filippenko}, Alexei V. and {Challis}, Peter and {Clocchiatti}, Alejandro and {Diercks}, Alan and {Garnavich}, Peter M. and {Gilliland}, Ron L. and {Hogan}, Craig J. and {Jha}, Saurabh and {Kirshner}, Robert P. and {Leibundgut}, B. and {Phillips}, M.~M. and {Reiss}, David and {Schmidt}, Brian P. and {Schommer}, Robert A. and {Smith}, R. Chris and {Spyromilio}, J. and {Stubbs}, Christopher and {Suntzeff}, Nicholas B. and {Tonry}, John},
        title = "{Observational Evidence from Supernovae for an Accelerating Universe and a Cosmological Constant}",
      journal = {\aj},
         year = 1998,
        month = sep,
       volume = {116},
       number = {3},
        pages = {1009-1038},
          doi = {10.1086/300499},
archivePrefix = {arXiv},
       eprint = {astro-ph/9805201},
 primaryClass = {astro-ph},
       adsurl = {https://ui.adsabs.harvard.edu/abs/1998AJ....116.1009R}
}

@ARTICLE{1999ApJ...517..565P,
       author = {{Perlmutter}, S. and {Aldering}, G. and {Goldhaber}, G. and {Knop}, R.~A. and {Nugent}, P. and {Castro}, P.~G. and {Deustua}, S. and {Fabbro}, S. and {Goobar}, A. and {Groom}, D.~E. and {Hook}, I.~M. and {Kim}, A.~G. and {Kim}, M.~Y. and {Lee}, J.~C. and {Nunes}, N.~J. and {Pain}, R. and {Pennypacker}, C.~R. and {Quimby}, R. and {Lidman}, C. and {Ellis}, R.~S. and {Irwin}, M. and {McMahon}, R.~G. and {Ruiz-Lapuente}, P. and {Walton}, N. and {Schaefer}, B. and {Boyle}, B.~J. and {Filippenko}, A.~V. and {Matheson}, T. and {Fruchter}, A.~S. and {Panagia}, N. and {Newberg}, H.~J.~M. and {Couch}, W.~J. and {Project}, The Supernova Cosmology},
        title = "{Measurements of {\ensuremath{\Omega}} and {\ensuremath{\Lambda}} from 42 High-Redshift Supernovae}",
      journal = {\apj},
         year = 1999,
        month = jun,
       volume = {517},
       number = {2},
        pages = {565-586},
          doi = {10.1086/307221},
archivePrefix = {arXiv},
       eprint = {astro-ph/9812133},
 primaryClass = {astro-ph},
       adsurl = {https://ui.adsabs.harvard.edu/abs/1999ApJ...517..565P}
}

@ARTICLE{2013PhR...530...87W,
       author = {{Weinberg}, David H. and {Mortonson}, Michael J. and {Eisenstein}, Daniel J. and {Hirata}, Christopher and {Riess}, Adam G. and {Rozo}, Eduardo},
        title = "{Observational probes of cosmic acceleration}",
      journal = {\physrep},
         year = 2013,
        month = sep,
       volume = {530},
       number = {2},
        pages = {87-255},
          doi = {10.1016/j.physrep.2013.05.001},
archivePrefix = {arXiv},
       eprint = {1201.2434},
 primaryClass = {astro-ph.CO},
       adsurl = {https://ui.adsabs.harvard.edu/abs/2013PhR...530...87W}
}

@ARTICLE{2022NewAR..9501659P,
       author = {{Perivolaropoulos}, L. and {Skara}, F.},
        title = "{Challenges for {\ensuremath{\Lambda}}CDM: An update}",
      journal = {\nar},
         year = 2022,
        month = dec,
       volume = {95},
          eid = {101659},
        pages = {101659},
          doi = {10.1016/j.newar.2022.101659},
archivePrefix = {arXiv},
       eprint = {2105.05208},
 primaryClass = {astro-ph.CO},
       adsurl = {https://ui.adsabs.harvard.edu/abs/2022NewAR..9501659P}
}

@ARTICLE{2006JCAP...11..003R,
       author = {{R{\"a}s{\"a}nen}, Syksy},
        title = "{Accelerated expansion from structure formation}",
      journal = {\jcap},
         year = 2006,
        month = nov,
       volume = {2006},
       number = {11},
          eid = {003},
        pages = {003},
          doi = {10.1088/1475-7516/2006/11/003},
archivePrefix = {arXiv},
       eprint = {astro-ph/0607626},
 primaryClass = {astro-ph},
       adsurl = {https://ui.adsabs.harvard.edu/abs/2006JCAP...11..003R}
}

@ARTICLE{2008GReGr..40..451E,
       author = {{Enqvist}, Kari},
        title = "{Lemaitre Tolman Bondi model and accelerating expansion}",
      journal = {General Relativity and Gravitation},
         year = 2008,
        month = feb,
       volume = {40},
       number = {2-3},
        pages = {451-466},
          doi = {10.1007/s10714-007-0553-9},
archivePrefix = {arXiv},
       eprint = {0709.2044},
 primaryClass = {astro-ph},
       adsurl = {https://ui.adsabs.harvard.edu/abs/2008GReGr..40..451E}
}

@ARTICLE{2007CQGra..24.5985C,
       author = {{Catto{\"e}n}, C{\'e}line and {Visser}, Matt},
        title = "{The Hubble series: convergence properties and redshift variables}",
      journal = {Classical and Quantum Gravity},
         year = 2007,
        month = dec,
       volume = {24},
       number = {23},
        pages = {5985-5997},
          doi = {10.1088/0264-9381/24/23/018},
archivePrefix = {arXiv},
       eprint = {0710.1887},
 primaryClass = {gr-qc},
       adsurl = {https://ui.adsabs.harvard.edu/abs/2007CQGra..24.5985C}
}

@ARTICLE{2020MNRAS.494.2576C,
       author = {{Capozziello}, S. and {D'Agostino}, R. and {Luongo}, O.},
        title = "{High-redshift cosmography: auxiliary variables versus Pad{\'e} polynomials}",
      journal = {\mnras},
         year = 2020,
        month = may,
       volume = {494},
       number = {2},
        pages = {2576-2590},
          doi = {10.1093/mnras/staa871},
archivePrefix = {arXiv},
       eprint = {2003.09341},
 primaryClass = {astro-ph.CO},
       adsurl = {https://ui.adsabs.harvard.edu/abs/2020MNRAS.494.2576C}
}

@ARTICLE{2022AA...661A..71H,
       author = {{Hu}, J.~P. and {Wang}, F.~Y.},
        title = "{High-redshift cosmography: Application and comparison with different methods}",
      journal = {\aap},
         year = 2022,
        month = may,
       volume = {661},
          eid = {A71},
        pages = {A71},
          doi = {10.1051/0004-6361/202142162},
archivePrefix = {arXiv},
       eprint = {2202.09075},
 primaryClass = {astro-ph.CO},
       adsurl = {https://ui.adsabs.harvard.edu/abs/2022A&A...661A..71H}
}

@ARTICLE{2024AA...689A.215H,
       author = {{Hu}, J.~P. and {Hu}, J. and {Jia}, X.~D. and {Gao}, B.~Q. and {Wang}, F.~Y.},
        title = "{Testing cosmic anisotropy with Pad{\'e} approximations and the latest Pantheon+ sample}",
      journal = {\aap},
         year = 2024,
        month = sep,
       volume = {689},
          eid = {A215},
        pages = {A215},
          doi = {10.1051/0004-6361/202450342},
archivePrefix = {arXiv},
       eprint = {2406.14827},
 primaryClass = {astro-ph.CO},
       adsurl = {https://ui.adsabs.harvard.edu/abs/2024A&A...689A.215H}
}

@ARTICLE{2022ApJ...938..110B,
       author = {{Brout}, Dillon and {Scolnic}, Dan and {Popovic}, Brodie and {Riess}, Adam G. and {Carr}, Anthony and {Zuntz}, Joe and {Kessler}, Rick and {Davis}, Tamara M. and {Hinton}, Samuel and {Jones}, David and {Kenworthy}, W. D'Arcy and {Peterson}, Erik R. and {Said}, Khaled and {Taylor}, Georgie and {Ali}, Noor and {Armstrong}, Patrick and {Charvu}, Pranav and {Dwomoh}, Arianna and {Meldorf}, Cole and {Palmese}, Antonella and {Qu}, Helen and {Rose}, Benjamin M. and {Sanchez}, Bruno and {Stubbs}, Christopher W. and {Vincenzi}, Maria and {Wood}, Charlotte M. and {Brown}, Peter J. and {Chen}, Rebecca and {Chambers}, Ken and {Coulter}, David A. and {Dai}, Mi and {Dimitriadis}, Georgios and {Filippenko}, Alexei V. and {Foley}, Ryan J. and {Jha}, Saurabh W. and {Kelsey}, Lisa and {Kirshner}, Robert P. and {M{\"o}ller}, Anais and {Muir}, Jessie and {Nadathur}, Seshadri and {Pan}, Yen-Chen and {Rest}, Armin and {Rojas-Bravo}, Cesar and {Sako}, Masao and {Siebert}, Matthew R. and {Smith}, Mat and {Stahl}, Benjamin E. and {Wiseman}, Phil},
        title = "{The Pantheon+ Analysis: Cosmological Constraints}",
      journal = {\apj},
         year = 2022,
        month = oct,
       volume = {938},
       number = {2},
          eid = {110},
        pages = {110},
          doi = {10.3847/1538-4357/ac8e04},
archivePrefix = {arXiv},
       eprint = {2202.04077},
 primaryClass = {astro-ph.CO},
       adsurl = {https://ui.adsabs.harvard.edu/abs/2022ApJ...938..110B}
}

@BOOK{Baker1996,
       author = {{Baker}, George A. and {Graves-Morris}, Peter},
        title = "{Pade approximants}",
         year = 1996,
    publisher = {Cambridge University Press},
      address = {Cambridge, UK}
}

@ARTICLE{2018ApJ...859..101S,
       author = {{Scolnic}, D.~M. and {Jones}, D.~O. and {Rest}, A. and {Pan}, Y.~C. and {Chornock}, R. and {Foley}, R.~J. and {Huber}, M.~E. and {Kessler}, R. and {Narayan}, G. and {Riess}, A.~G. and {Rodney}, S. and {Berger}, E. and {Brout}, D.~J. and {Challis}, P.~J. and {Drout}, M. and {Finkbeiner}, D. and {Lunnan}, R. and {Kirshner}, R.~P. and {Sanders}, N.~E. and {Schlafly}, E. and {Smartt}, S. and {Stubbs}, C.~W. and {Tonry}, J. and {Wood-Vasey}, W.~M. and {Foley}, M. and {Hand}, J. and {Johnson}, E. and {Burgett}, W.~S. and {Chambers}, K.~C. and {Draper}, P.~W. and {Hodapp}, K.~W. and {Kaiser}, N. and {Kudritzki}, R.~P. and {Magnier}, E.~A. and {Metcalfe}, N. and {Bresolin}, F. and {Gall}, E. and {Kotak}, R. and {McCrum}, M. and {Smith}, K.~W.},
        title = "{The Complete Light-curve Sample of Spectroscopically Confirmed SNe Ia from Pan-STARRS1 and Cosmological Constraints from the Combined Pantheon Sample}",
      journal = {\apj},
         year = 2018,
        month = jun,
       volume = {859},
       number = {2},
          eid = {101},
        pages = {101},
          doi = {10.3847/1538-4357/aab9bb},
archivePrefix = {arXiv},
       eprint = {1710.00845},
 primaryClass = {astro-ph.CO},
       adsurl = {https://ui.adsabs.harvard.edu/abs/2018ApJ...859..101S}
}

@ARTICLE{2013PASP..125..306F,
       author = {{Foreman-Mackey}, Daniel and {Hogg}, David W. and {Lang}, Dustin and {Goodman}, Jonathan},
        title = "{emcee: The MCMC Hammer}",
      journal = {\pasp},
         year = 2013,
        month = mar,
       volume = {125},
       number = {925},
        pages = {306},
          doi = {10.1086/670067},
archivePrefix = {arXiv},
       eprint = {1202.3665},
 primaryClass = {astro-ph.IM},
       adsurl = {https://ui.adsabs.harvard.edu/abs/2013PASP..125..306F}
}

@ARTICLE{1978AnSta...6..461S,
       author = {{Schwarz}, Gideon},
        title = "{Estimating the Dimension of a Model}",
      journal = {Annals of Statistics},
         year = 1978,
        month = jul,
       volume = {6},
       number = {2},
        pages = {461-464},
       adsurl = {https://ui.adsabs.harvard.edu/abs/1978AnSta...6..461S}
}

@ARTICLE{2019MNRAS.488.5728E,
       author = {{Earp}, Samuel W.~F. and {Debattista}, Victor P. and {Macci{\`o}}, Andrea V. and {Wang}, Liang and {Buck}, Tobias and {Khachaturyants}, Tigran},
        title = "{Drivers of disc tilting I: correlations and possible drivers for Milky Way analogues}",
      journal = {\mnras},
         year = 2019,
        month = oct,
       volume = {488},
       number = {4},
        pages = {5728-5738},
          doi = {10.1093/mnras/stz2109},
archivePrefix = {arXiv},
       eprint = {1907.10969},
 primaryClass = {astro-ph.GA},
       adsurl = {https://ui.adsabs.harvard.edu/abs/2019MNRAS.488.5728E}
}

@ARTICLE{2022ApJ...934L...7R,
       author = {{Riess}, Adam G. and {Yuan}, Wenlong and {Macri}, Lucas M. and {Scolnic}, Dan and {Brout}, Dillon and {Casertano}, Stefano and {Jones}, David O. and {Murakami}, Yukei and {Anand}, Gagandeep S. and {Breuval}, Louise and {Brink}, Thomas G. and {Filippenko}, Alexei V. and {Hoffmann}, Samantha and {Jha}, Saurabh W. and {D'arcy Kenworthy}, W. and {Mackenty}, John and {Stahl}, Benjamin E. and {Zheng}, WeiKang},
        title = "{A Comprehensive Measurement of the Local Value of the Hubble Constant with 1 km s$^{-1}$ Mpc$^{-1}$ Uncertainty from the Hubble Space Telescope and the SH0ES Team}",
      journal = {\apjl},
         year = 2022,
        month = jul,
       volume = {934},
       number = {1},
          eid = {L7},
        pages = {L7},
          doi = {10.3847/2041-8213/ac5c5b},
archivePrefix = {arXiv},
       eprint = {2112.04510},
 primaryClass = {astro-ph.CO},
       adsurl = {https://ui.adsabs.harvard.edu/abs/2022ApJ...934L...7R}
}

@ARTICLE{2006PhRvD..74j3518L,
       author = {{Linder}, Eric V.},
        title = "{Importance of supernovae at z&lt;0.1 for probing dark energy}",
      journal = {\prd},
         year = 2006,
        month = nov,
       volume = {74},
       number = {10},
          eid = {103518},
        pages = {103518},
          doi = {10.1103/PhysRevD.74.103518},
archivePrefix = {arXiv},
       eprint = {astro-ph/0609507},
 primaryClass = {astro-ph},
       adsurl = {https://ui.adsabs.harvard.edu/abs/2006PhRvD..74j3518L}
}

@article{Sugiura1978,
  author  = {{Sugiura}, N.},
    title = "{Further analysts of the data by akaike' s information criterion and the finite corrections}",
  journal = {Commun. Stat. Theory Methods},
     year = {1978},
   volume = {7},
   number = {1},
    pages = {13-26},
      doi = {10.1080/03610927808827599},
      url = {http://dx.doi.org/10.1080/03610927808827599},
   eprint = {http://dx.doi.org/10.1080/03610927808827599}
}

@ARTICLE{2024ApJ...971...19C,
       author = {{Clocchiatti}, Alejandro and {Rodr{\'\i}guez}, {\'O}smar and {{\'O}rdenes Morales}, Ariel and {Cuevas-Tapia}, Benjamin},
        title = "{Global Anisotropies of {\ensuremath{\Omega}}$_{{\ensuremath{\Lambda}}}$}",
      journal = {\apj},
         year = 2024,
        month = aug,
       volume = {971},
       number = {1},
          eid = {19},
        pages = {19},
          doi = {10.3847/1538-4357/ad51ff},
archivePrefix = {arXiv},
       eprint = {2406.00273},
 primaryClass = {astro-ph.CO},
       adsurl = {https://ui.adsabs.harvard.edu/abs/2024ApJ...971...19C}
}

@ARTICLE{2024AA...681A..88H,
       author = {{Hu}, J.~P. and {Wang}, Y.~Y. and {Hu}, J. and {Wang}, F.~Y.},
        title = "{Testing the cosmological principle with the Pantheon+ sample and the region-fitting method}",
      journal = {\aap},
         year = 2024,
        month = jan,
       volume = {681},
          eid = {A88},
        pages = {A88},
          doi = {10.1051/0004-6361/202347121},
archivePrefix = {arXiv},
       eprint = {2310.11727},
 primaryClass = {astro-ph.CO},
       adsurl = {https://ui.adsabs.harvard.edu/abs/2024A&A...681A..88H}
}

@ARTICLE{2015ApJ...814....7B,
       author = {{Bochner}, Brett and {Pappas}, Damon and {Dong}, Menglu},
        title = "{Testing Lambda and the Limits of Cosmography with the Union2.1 Supernova Compilation}",
      journal = {\apj},
         year = 2015,
        month = nov,
       volume = {814},
       number = {1},
          eid = {7},
        pages = {7},
          doi = {10.1088/0004-637X/814/1/7},
archivePrefix = {arXiv},
       eprint = {1308.6050},
 primaryClass = {astro-ph.CO},
       adsurl = {https://ui.adsabs.harvard.edu/abs/2015ApJ...814....7B}
}

@ARTICLE{2024ApJ...971L..32P,
       author = {{Pierel}, J.~D.~R. and {Engesser}, M. and {Coulter}, D.~A. and {DeCoursey}, C. and {Siebert}, M.~R. and {Rest}, A. and {Egami}, E. and {Chen}, W. and {Fox}, O.~D. and {Jones}, D.~O. and {Joshi}, B.~A. and {Moriya}, T.~J. and {Zenati}, Y. and {Bunker}, A.~J. and {Cargile}, P.~A. and {Curti}, M. and {Eisenstein}, D.~J. and {Gezari}, S. and {Gomez}, S. and {Guolo}, M. and {Johnson}, B.~D. and {Karmen}, M. and {Maiolino}, R. and {Quimby}, R.~M. and {Robertson}, B. and {Shahbandeh}, M. and {Strolger}, L.~G. and {Sun}, F. and {Wang}, Q. and {Wevers}, T.},
        title = "{Discovery of an Apparent Red, High-velocity Type Ia Supernova at z = 2.9 with JWST}",
      journal = {\apjl},
         year = 2024,
        month = aug,
       volume = {971},
       number = {2},
          eid = {L32},
        pages = {L32},
          doi = {10.3847/2041-8213/ad6908},
archivePrefix = {arXiv},
       eprint = {2406.05089},
 primaryClass = {astro-ph.GA},
       adsurl = {https://ui.adsabs.harvard.edu/abs/2024ApJ...971L..32P}
}

@ARTICLE{2025ApJ...981L...9P,
       author = {{Pierel}, J.~D.~R. and {Coulter}, D.~A. and {Siebert}, M.~R. and {Akins}, H.~B. and {Engesser}, M. and {Fox}, O.~D. and {Franco}, M. and {Rest}, A. and {Agrawal}, A. and {Ajay}, Y. and {Allen}, N. and {Casey}, C.~M. and {DeCoursey}, C. and {Drakos}, N.~E. and {Egami}, E. and {Faisst}, A.~L. and {Gezari}, S. and {Gozaliasl}, G. and {Ilbert}, O. and {Jones}, D.~O. and {Karmen}, M. and {Kartaltepe}, J.~S. and {Koekemoer}, A.~M. and {Lane}, Z.~G. and {Larson}, R.~L. and {Li}, T. and {Liu}, D. and {Moriya}, T.~J. and {McCracken}, H.~J. and {Paquereau}, L. and {Quimby}, R.~M. and {Rich}, R.~M. and {Rhodes}, J. and {Robertson}, B.~E. and {Sanders}, D.~B. and {Shahbandeh}, M. and {Shuntov}, M. and {Silverman}, J.~D. and {Strolger}, L.~G. and {Toft}, S. and {Zenati}, Y.},
        title = "{Testing for Intrinsic Type Ia Supernova Luminosity Evolution at z &gt; 2 with JWST}",
      journal = {\apjl},
         year = 2025,
        month = mar,
       volume = {981},
       number = {1},
          eid = {L9},
        pages = {L9},
          doi = {10.3847/2041-8213/adb1d9},
archivePrefix = {arXiv},
       eprint = {2411.11953},
 primaryClass = {astro-ph.CO},
       adsurl = {https://ui.adsabs.harvard.edu/abs/2025ApJ...981L...9P}
}

@ARTICLE{2025ApJ...979..250D,
       author = {{DeCoursey}, Christa and {Egami}, Eiichi and {Pierel}, Justin D.~R. and {Sun}, Fengwu and {Rest}, Armin and {Coulter}, David A. and {Engesser}, Michael and {Siebert}, Matthew R. and {Hainline}, Kevin N. and {Johnson}, Benjamin D. and {Bunker}, Andrew J. and {Cargile}, Phillip A. and {Charlot}, Stephane and {Chen}, Wenlei and {Curti}, Mirko and {DeFour-Remy}, Shea and {Eisenstein}, Daniel J. and {Fox}, Ori D. and {Gezari}, Suvi and {Gomez}, Sebastian and {Jencson}, Jacob and {Joshi}, Bhavin A. and {Khairnar}, Sanvi and {Lyu}, Jianwei and {Maiolino}, Roberto and {Moriya}, Takashi J. and {Quimby}, Robert M. and {Rieke}, George H. and {Rieke}, Marcia J. and {Robertson}, Brant and {Shahbandeh}, Melissa and {Strolger}, Louis-Gregory and {Tacchella}, Sandro and {Wang}, Qinan and {Williams}, Christina C. and {Willmer}, Christopher N.~A. and {Willott}, Chris and {Zenati}, Yossef},
        title = "{The JADES Transient Survey: Discovery and Classification of Supernovae in the JADES Deep Field}",
      journal = {\apj},
         year = 2025,
        month = feb,
       volume = {979},
       number = {2},
          eid = {250},
        pages = {250},
          doi = {10.3847/1538-4357/ad8fab},
archivePrefix = {arXiv},
       eprint = {2406.05060},
 primaryClass = {astro-ph.HE},
       adsurl = {https://ui.adsabs.harvard.edu/abs/2025ApJ...979..250D}
}

@ARTICLE{2023arXiv230702670H,
       author = {{Hounsell}, Rebekah and {Scolnic}, Dan and {Brout}, Dillon and {Rose}, Benjamin and {Fox}, Ori and {Sako}, Masao and {Macias}, Phillip and {Joshi}, Bhavin and {Desutua}, Susana and {Rubin}, David and {Casertano}, Stefano and {Perlmutter}, Saul and {Aldering}, Greg and {Mandel}, Kaisey and {Sosey}, Megan and {Suzuki}, Nao and {Ryan}, Russell},
        title = "{Roman CCS White Paper: Measuring Type Ia Supernovae Discovered in the Roman High Latitude Time Domain Survey}",
      journal = {arXiv e-prints},
         year = 2023,
        month = jul,
          eid = {arXiv:2307.02670},
        pages = {arXiv:2307.02670},
          doi = {10.48550/arXiv.2307.02670},
archivePrefix = {arXiv},
       eprint = {2307.02670},
 primaryClass = {astro-ph.IM},
       adsurl = {https://ui.adsabs.harvard.edu/abs/2023arXiv230702670H}
}

@ARTICLE{2023SCPMA..6629511L,
       author = {{Li}, Shi-Yu and {Li}, Yun-Long and {Zhang}, Tianmeng and {Vink{\'o}}, J{\'o}zsef and {Reg{\H{o}}s}, Enik{\H{o}} and {Wang}, Xiaofeng and {Xi}, Gaobo and {Zhan}, Hu},
        title = "{Forecast of cosmological constraints with type Ia supernovae from the Chinese Space Station Telescope}",
      journal = {Science China Physics, Mechanics, and Astronomy},
         year = 2023,
        month = feb,
       volume = {66},
       number = {2},
          eid = {229511},
        pages = {229511},
          doi = {10.1007/s11433-022-2018-0},
archivePrefix = {arXiv},
       eprint = {2210.05450},
 primaryClass = {astro-ph.CO},
       adsurl = {https://ui.adsabs.harvard.edu/abs/2023SCPMA..6629511L}
}

@ARTICLE{2007AA...474..717S,
       author = {{Schwarz}, D.~J. and {Weinhorst}, B.},
        title = "{(An)isotropy of the Hubble diagram: comparing hemispheres}",
      journal = {\aap},
         year = 2007,
        month = nov,
       volume = {474},
       number = {3},
        pages = {717-729},
          doi = {10.1051/0004-6361:20077998},
archivePrefix = {arXiv},
       eprint = {0706.0165},
 primaryClass = {astro-ph},
       adsurl = {https://ui.adsabs.harvard.edu/abs/2007A&A...474..717S}
}

@ARTICLE{2023PhRvD.108l3533M,
       author = {{Mc Conville}, Ruair{\'\i} and {{\'O} Colg{\'a}in}, Eoin},
        title = "{Anisotropic distance ladder in Pantheon+supernovae}",
      journal = {\prd},
         year = 2023,
        month = dec,
       volume = {108},
       number = {12},
          eid = {123533},
        pages = {123533},
          doi = {10.1103/PhysRevD.108.123533},
archivePrefix = {arXiv},
       eprint = {2304.02718},
 primaryClass = {astro-ph.CO},
       adsurl = {https://ui.adsabs.harvard.edu/abs/2023PhRvD.108l3533M}
}

@ARTICLE{2024MNRAS.533.2615C,
       author = {{Camilleri}, R. and {Davis}, T.~M. and {Vincenzi}, M. and {Shah}, P. and {Frieman}, J. and {Kessler}, R. and {Armstrong}, P. and {Brout}, D. and {Carr}, A. and {Chen}, R. and {Galbany}, L. and {Glazebrook}, K. and {Hinton}, S.~R. and {Lee}, J. and {Lidman}, C. and {M{\"o}ller}, A. and {Popovic}, B. and {Qu}, H. and {Sako}, M. and {Scolnic}, D. and {Smith}, M. and {Sullivan}, M. and {S{\'a}nchez}, B.~O. and {Taylor}, G. and {Toy}, M. and {Wiseman}, P. and {Abbott}, T.~M.~C. and {Aguena}, M. and {Allam}, S. and {Alves}, O. and {Annis}, J. and {Avila}, S. and {Bacon}, D. and {Bertin}, E. and {Bocquet}, S. and {Brooks}, D. and {Burke}, D.~L. and {Carnero Rosell}, A. and {Carretero}, J. and {Castander}, F.~J. and {da Costa}, L.~N. and {Pereira}, M.~E.~S. and {Desai}, S. and {Diehl}, H.~T. and {Doel}, P. and {Doux}, C. and {Everett}, S. and {Ferrero}, I. and {Flaugher}, B. and {Fosalba}, P. and {Garc{\'\i}a-Bellido}, J. and {Gatti}, M. and {Gaztanaga}, E. and {Giannini}, G. and {Gruen}, D. and {Hollowood}, D.~L. and {Honscheid}, K. and {James}, D.~J. and {Kuehn}, K. and {Lahav}, O. and {Lee}, S. and {Lewis}, G.~F. and {Marshall}, J.~L. and {Mena-Fern{\'a}ndez}, J. and {Miquel}, R. and {Muir}, J. and {Myles}, J. and {Ogando}, R.~L.~C. and {Pieres}, A. and {Malag{\'o}n}, A.~A. Plazas and {Porredon}, A. and {Rodriguez-Monroy}, M. and {Sanchez}, E. and {Sanchez Cid}, D. and {Schubnell}, M. and {Sevilla-Noarbe}, I. and {Suchyta}, E. and {Swanson}, M.~E.~C. and {Tarle}, G. and {Walker}, A.~R. and {Weaverdyck}, N. and {DES Collaboration}},
        title = "{The dark energy survey supernova program: investigating beyond-{\ensuremath{\Lambda}}CDM}",
      journal = {\mnras},
         year = 2024,
        month = sep,
       volume = {533},
       number = {3},
        pages = {2615-2639},
          doi = {10.1093/mnras/stae1988},
archivePrefix = {arXiv},
       eprint = {2406.05048},
 primaryClass = {astro-ph.CO},
       adsurl = {https://ui.adsabs.harvard.edu/abs/2024MNRAS.533.2615C}
}

@ARTICLE{2025MNRAS.541.2585V,
       author = {{Vincenzi}, M. and {Kessler}, R. and {Shah}, P. and {Lee}, J. and {Davis}, T.~M. and {Scolnic}, D. and {Armstrong}, P. and {Brout}, D. and {Camilleri}, R. and {Chen}, R. and {Galbany}, L. and {Lidman}, C. and {M{\"o}ller}, A. and {Popovic}, B. and {Rose}, B. and {Sako}, M. and {S{\'a}nchez}, B.~O. and {Smith}, M. and {Sullivan}, M. and {Wiseman}, P. and {Abbott}, T.~M.~C. and {Aguena}, M. and {Allam}, S. and {Andrade-Oliveira}, F. and {Bocquet}, S. and {Brooks}, D. and {Carnero Rosell}, A. and {Carretero}, J. and {da Costa}, L.~N. and {Pereira}, M.~E.~S. and {Diehl}, H.~T. and {Doel}, P. and {Everett}, S. and {Flaugher}, B. and {Frieman}, J. and {Garc{\'\i}a-Bellido}, J. and {Gaztanaga}, E. and {Gruen}, D. and {Gruendl}, R.~A. and {Gutierrez}, G. and {Hinton}, S.~R. and {Hollowood}, D.~L. and {Honscheid}, K. and {James}, D.~J. and {Kuehn}, K. and {Lahav}, O. and {Lee}, S. and {Marshall}, J.~L. and {Mena-Fern{\'a}ndez}, J. and {Miquel}, R. and {Muir}, J. and {Myles}, J. and {Palmese}, A. and {Plazas Malag{\'o}n}, A.~A. and {Porredon}, A. and {Samuroff}, S. and {Sanchez}, E. and {Sanchez Cid}, D. and {Sevilla-Noarbe}, I. and {Suchyta}, E. and {Tarle}, G. and {To}, C. and {Tucker}, D.~L. and {Vikram}, V. and {Walker}, A.~R. and {Weaverdyck}, N. and {Weller}, J.},
        title = "{Comparing the DES-SN5YR and Pantheon+ SN cosmology analyses: investigation based on 'evolving dark energy or supernovae systematics'?}",
      journal = {\mnras},
         year = 2025,
        month = aug,
       volume = {541},
       number = {3},
        pages = {2585-2593},
          doi = {10.1093/mnras/staf943},
archivePrefix = {arXiv},
       eprint = {2501.06664},
 primaryClass = {astro-ph.CO},
       adsurl = {https://ui.adsabs.harvard.edu/abs/2025MNRAS.541.2585V}
}

@ARTICLE{2024MNRAS.529.2100P,
       author = {{Popovic}, Brodie and {Scolnic}, Daniel and {Vincenzi}, Maria and {Sullivan}, Mark and {Brout}, Dillon and {Chen}, Rebecca and {Patel}, Utsav and {Peterson}, Erik R. and {Kessler}, Richard and {Kelsey}, Lisa and {Sanchez}, Bruno O. and {Bailey}, Ava Claire and {Wiseman}, Phil and {Toy}, Marcus},
        title = "{Amalgame: cosmological constraints from the first combined photometric supernova sample}",
      journal = {\mnras},
         year = 2024,
        month = apr,
       volume = {529},
       number = {3},
        pages = {2100-2115},
          doi = {10.1093/mnras/stae420},
archivePrefix = {arXiv},
       eprint = {2309.05654},
 primaryClass = {astro-ph.CO},
       adsurl = {https://ui.adsabs.harvard.edu/abs/2024MNRAS.529.2100P}
}

@ARTICLE{2011ApJ...738..162S,
       author = {{Sako}, Masao and {Bassett}, Bruce and {Connolly}, Brian and {Dilday}, Benjamin and {Cambell}, Heather and {Frieman}, Joshua A. and {Gladney}, Larry and {Kessler}, Richard and {Lampeitl}, Hubert and {Marriner}, John and {Miquel}, Ramon and {Nichol}, Robert C. and {Schneider}, Donald P. and {Smith}, Mathew and {Sollerman}, Jesper},
        title = "{Photometric Type Ia Supernova Candidates from the Three-year SDSS-II SN Survey Data}",
      journal = {\apj},
         year = 2011,
        month = sep,
       volume = {738},
       number = {2},
          eid = {162},
        pages = {162},
          doi = {10.1088/0004-637X/738/2/162},
archivePrefix = {arXiv},
       eprint = {1107.5106},
 primaryClass = {astro-ph.CO},
       adsurl = {https://ui.adsabs.harvard.edu/abs/2011ApJ...738..162S}
}

@ARTICLE{2026JHEAp..4900444A,
       author = {{Alfano}, Anna Chiara and {Cafaro}, Carlo and {Capozziello}, Salvatore and {Luongo}, Orlando and {Muccino}, Marco},
        title = "{Investigating the cosmic distance duality relation with gamma-ray bursts}",
      journal = {Journal of High Energy Astrophysics},
         year = 2026,
        month = jan,
       volume = {49},
          eid = {100444},
        pages = {100444},
          doi = {10.1016/j.jheap.2025.100444},
archivePrefix = {arXiv},
       eprint = {2509.09247},
 primaryClass = {astro-ph.CO},
       adsurl = {https://ui.adsabs.harvard.edu/abs/2026JHEAp..4900444A}
}

@ARTICLE{2025PhRvD.112h3515A,
       author = {{Abdul Karim}, M. and {Aguilar}, J. and {Ahlen}, S. and {Alam}, S. and {Allen}, L. and {Prieto}, C. Allende and {Alves}, O. and {Anand}, A. and {Andrade}, U. and {Armengaud}, E. and {Aviles}, A. and {Bailey}, S. and {Baltay}, C. and {Bansal}, P. and {Bault}, A. and {Behera}, J. and {BenZvi}, S. and {Bianchi}, D. and {Blake}, C. and {Brieden}, S. and {Brodzeller}, A. and {Brooks}, D. and {Buckley-Geer}, E. and {Burtin}, E. and {Calderon}, R. and {Canning}, R. and {Rosell}, A. Carnero and {Carrilho}, P. and {Casas}, L. and {Castander}, F.~J. and {Charles}, M. and {Chaussidon}, E. and {Chaves-Montero}, J. and {Chebat}, D. and {Chen}, X. and {Claybaugh}, T. and {Cole}, S. and {Cooper}, A.~P. and {Cuceu}, A. and {Dawson}, K.~S. and {de la Macorra}, A. and {de Mattia}, A. and {Deiosso}, N. and {Della Costa}, J. and {Demina}, R. and {Dey}, A. and {Dey}, B. and {Ding}, Z. and {Doel}, P. and {Edelstein}, J. and {Eisenstein}, D.~J. and {Elbers}, W. and {Fagrelius}, P. and {Fanning}, K. and {Fern{\'a}ndez-Garc{\'\i}a}, E. and {Ferraro}, S. and {Font-Ribera}, A. and {Forero-Romero}, J.~E. and {Frenk}, C.~S. and {Garcia-Quintero}, C. and {Garrison}, L.~H. and {Gazta{\~n}aga}, E. and {Gil-Mar{\'\i}n}, H. and {Gontcho A Gontcho}, S. and {Gonzalez}, D. and {Gonzalez-Morales}, A.~X. and {Gordon}, C. and {Green}, D. and {Gutierrez}, G. and {Guy}, J. and {Hadzhiyska}, B. and {Hahn}, C. and {He}, S. and {Herbold}, M. and {Herrera-Alcantar}, H.~K. and {Ho}, M.-F. and {Honscheid}, K. and {Howlett}, C. and {Huterer}, D. and {Ishak}, M. and {Juneau}, S. and {Kamble}, N.~V. and {Kara{\c{c}}ayl{\i}}, N.~G. and {Kehoe}, R. and {Kent}, S. and {Kim}, A.~G. and {Kirkby}, D. and {Kisner}, T. and {Koposov}, S.~E. and {Kremin}, A. and {Krolewski}, A. and {Lahav}, O. and {Lamman}, C. and {Landriau}, M. and {Lang}, D. and {Lasker}, J. and {Le Goff}, J.~M. and {Le Guillou}, L. and {Leauthaud}, A. and {Levi}, M.~E. and {Li}, Q. and {Li}, T.~S. and {Lodha}, K. and {Lokken}, M. and {Lozano-Rodr{\'\i}guez}, F. and {Magneville}, C. and {Manera}, M. and {Martini}, P. and {Matthewson}, W.~L. and {Meisner}, A. and {Mena-Fern{\'a}ndez}, J. and {Menegas}, A. and {Mergulh{\~a}o}, T. and {Miquel}, R. and {Moustakas}, J. and {Mu{\~n}oz-Guti{\'e}rrez}, A. and {Mu{\~n}oz-Santos}, D. and {Myers}, A.~D. and {Nadathur}, S. and {Naidoo}, K. and {Napolitano}, L. and {Newman}, J.~A. and {Niz}, G. and {Noriega}, H.~E. and {Paillas}, E. and {Palanque-Delabrouille}, N. and {Pan}, J. and {Peacock}, J.~A. and {Pellejero Ibanez}, M. and {Percival}, W.~J. and {P{\'e}rez-Fern{\'a}ndez}, A. and {P{\'e}rez-R{\`a}fols}, I. and {Pieri}, M.~M. and {Poppett}, C. and {Prada}, F. and {Rabinowitz}, D. and {Raichoor}, A. and {Ram{\'\i}rez-P{\'e}rez}, C. and {Rashkovetskyi}, M. and {Ravoux}, C. and {Rich}, J. and {Rocher}, A. and {Rockosi}, C. and {Rohlf}, J. and {Rom{\'a}n-Herrera}, J.~O. and {Ross}, A.~J. and {Rossi}, G. and {Ruggeri}, R. and {Ruhlmann-Kleider}, V. and {Samushia}, L. and {Sanchez}, E. and {Sanders}, N. and {Schlegel}, D. and {Schubnell}, M. and {Seo}, H. and {Shafieloo}, A. and {Sharples}, R. and {Silber}, J. and {Sinigaglia}, F. and {Sprayberry}, D. and {Tan}, T. and {Tarl{\'e}}, G. and {Taylor}, P. and {Turner}, W. and {Ure{\~n}a-L{\'o}pez}, L.~A. and {Vaisakh}, R. and {Valdes}, F. and {Valogiannis}, G. and {Vargas-Maga{\~n}a}, M. and {Verde}, L. and {Walther}, M. and {Weaver}, B.~A. and {Weinberg}, D.~H. and {White}, M. and {Wolfson}, M. and {Y{\`e}che}, C. and {Yu}, J. and {Zaborowski}, E.~A. and {Zarrouk}, P. and {Zhai}, Z. and {Zhang}, H. and {Zhao}, C. and {Zhao}, G.~B. and {Zhou}, R. and {Zou}, H. and {DESI Collaboration}},
        title = "{DESI DR2 results. II. Measurements of baryon acoustic oscillations and cosmological constraints}",
      journal = {\prd},
         year = 2025,
        month = oct,
       volume = {112},
       number = {8},
          eid = {083515},
        pages = {083515},
          doi = {10.1103/tr6y-kpc6},
archivePrefix = {arXiv},
       eprint = {2503.14738},
 primaryClass = {astro-ph.CO},
       adsurl = {https://ui.adsabs.harvard.edu/abs/2025PhRvD.112h3515A}
}

@ARTICLE{2005ApJ...622..759G,
       author = {{G{\'o}rski}, K.~M. and {Hivon}, E. and {Banday}, A.~J. and {Wandelt}, B.~D. and {Hansen}, F.~K. and {Reinecke}, M. and {Bartelmann}, M.},
        title = "{HEALPix: A Framework for High-Resolution Discretization and Fast Analysis of Data Distributed on the Sphere}",
      journal = {\apj},
         year = 2005,
        month = apr,
       volume = {622},
       number = {2},
        pages = {759-771},
          doi = {10.1086/427976},
archivePrefix = {arXiv},
       eprint = {astro-ph/0409513},
 primaryClass = {astro-ph},
       adsurl = {https://ui.adsabs.harvard.edu/abs/2005ApJ...622..759G}
}

@ARTICLE{2019JOSS....4.1298Z,
       author = {{Zonca}, Andrea and {Singer}, Leo and {Lenz}, Daniel and {Reinecke}, Martin and {Rosset}, Cyrille and {Hivon}, Eric and {Gorski}, Krzysztof},
        title = "{healpy: equal area pixelization and spherical harmonics transforms for data on the sphere in Python}",
      journal = {The Journal of Open Source Software},
         year = 2019,
        month = mar,
       volume = {4},
       number = {35},
          eid = {1298},
        pages = {1298},
          doi = {10.21105/joss.01298},
       adsurl = {https://ui.adsabs.harvard.edu/abs/2019JOSS....4.1298Z}
}

@ARTICLE{2020JCAP...10..009B,
       author = {{Bayer}, Adrian E. and {Seljak}, Uro{\v{s}}},
        title = "{The look-elsewhere effect from a unified Bayesian and frequentist perspective}",
      journal = {\jcap},
         year = 2020,
        month = oct,
       volume = {2020},
       number = {10},
          eid = {009},
        pages = {009},
          doi = {10.1088/1475-7516/2020/10/009},
archivePrefix = {arXiv},
       eprint = {2007.13821},
 primaryClass = {physics.data-an},
       adsurl = {https://ui.adsabs.harvard.edu/abs/2020JCAP...10..009B}
}

@ARTICLE{2020MNRAS.491.4960L,
       author = {{Li}, En-Kun and {Du}, Minghui and {Xu}, Lixin},
        title = "{General cosmography model with spatial curvature}",
      journal = {\mnras},
         year = 2020,
        month = feb,
       volume = {491},
       number = {4},
        pages = {4960-4972},
          doi = {10.1093/mnras/stz3308},
archivePrefix = {arXiv},
       eprint = {1903.11433},
 primaryClass = {astro-ph.CO},
       adsurl = {https://ui.adsabs.harvard.edu/abs/2020MNRAS.491.4960L}
}

@ARTICLE{2025SCPMA..6800413H,
       author = {{Huang}, Lu and {Cai}, Rong-Gen and {Wang}, Shao-Jiang},
        title = "{The DESI DR1/DR2 evidence for dynamical dark energy is biased by low-redshift supernovae}",
      journal = {Science China Physics, Mechanics, and Astronomy},
         year = 2025,
        month = aug,
       volume = {68},
       number = {10},
          eid = {100413},
        pages = {100413},
          doi = {10.1007/s11433-025-2754-5},
archivePrefix = {arXiv},
       eprint = {2502.04212},
 primaryClass = {astro-ph.CO},
       adsurl = {https://ui.adsabs.harvard.edu/abs/2025SCPMA..6800413H}
}

@ARTICLE{2023ApJ...954...31C,
       author = {{Casey}, Caitlin M. and {Kartaltepe}, Jeyhan S. and {Drakos}, Nicole E. and {Franco}, Maximilien and {Harish}, Santosh and {Paquereau}, Louise and {Ilbert}, Olivier and {Rose}, Caitlin and {Cox}, Isabella G. and {Nightingale}, James W. and {Robertson}, Brant E. and {Silverman}, John D. and {Koekemoer}, Anton M. and {Massey}, Richard and {McCracken}, Henry Joy and {Rhodes}, Jason and {Akins}, Hollis B. and {Allen}, Natalie and {Amvrosiadis}, Aristeidis and {Arango-Toro}, Rafael C. and {Bagley}, Micaela B. and {Bongiorno}, Angela and {Capak}, Peter L. and {Champagne}, Jaclyn B. and {Chartab}, Nima and {Ch{\'a}vez Ortiz}, {\'O}scar A. and {Chworowsky}, Katherine and {Cooke}, Kevin C. and {Cooper}, Olivia R. and {Darvish}, Behnam and {Ding}, Xuheng and {Faisst}, Andreas L. and {Finkelstein}, Steven L. and {Fujimoto}, Seiji and {Gentile}, Fabrizio and {Gillman}, Steven and {Gould}, Katriona M.~L. and {Gozaliasl}, Ghassem and {Hayward}, Christopher C. and {He}, Qiuhan and {Hemmati}, Shoubaneh and {Hirschmann}, Michaela and {Jahnke}, Knud and {Jin}, Shuowen and {Khostovan}, Ali Ahmad and {Kokorev}, Vasily and {Lambrides}, Erini and {Laigle}, Clotilde and {Larson}, Rebecca L. and {Leung}, Gene C.~K. and {Liu}, Daizhong and {Liaudat}, Tobias and {Long}, Arianna S. and {Magdis}, Georgios and {Mahler}, Guillaume and {Mainieri}, Vincenzo and {Manning}, Sinclaire M. and {Maraston}, Claudia and {Martin}, Crystal L. and {McCleary}, Jacqueline E. and {McKinney}, Jed and {McPartland}, Conor J.~R. and {Mobasher}, Bahram and {Pattnaik}, Rohan and {Renzini}, Alvio and {Rich}, R. Michael and {Sanders}, David B. and {Sattari}, Zahra and {Scognamiglio}, Diana and {Scoville}, Nick and {Sheth}, Kartik and {Shuntov}, Marko and {Sparre}, Martin and {Suzuki}, Tomoko L. and {Talia}, Margherita and {Toft}, Sune and {Trakhtenbrot}, Benny and {Urry}, C. Megan and {Valentino}, Francesco and {Vanderhoof}, Brittany N. and {Vardoulaki}, Eleni and {Weaver}, John R. and {Whitaker}, Katherine E. and {Wilkins}, Stephen M. and {Yang}, Lilan and {Zavala}, Jorge A.},
        title = "{COSMOS-Web: An Overview of the JWST Cosmic Origins Survey}",
      journal = {\apj},
         year = 2023,
        month = sep,
       volume = {954},
       number = {1},
          eid = {31},
        pages = {31},
          doi = {10.3847/1538-4357/acc2bc},
archivePrefix = {arXiv},
       eprint = {2211.07865},
 primaryClass = {astro-ph.GA},
       adsurl = {https://ui.adsabs.harvard.edu/abs/2023ApJ...954...31C}
}

@ARTICLE{2024ApJ...975....5S,
       author = {{S{\'a}nchez}, B.~O. and {Brout}, D. and {Vincenzi}, M. and {Sako}, M. and {Herner}, K. and {Kessler}, R. and {Davis}, T.~M. and {Scolnic}, D. and {Acevedo}, M. and {Lee}, J. and {M{\"o}ller}, A. and {Qu}, H. and {Kelsey}, L. and {Wiseman}, P. and {Armstrong}, P. and {Rose}, B. and {Camilleri}, R. and {Chen}, R. and {Galbany}, L. and {Kovacs}, E. and {Lidman}, C. and {Popovic}, B. and {Smith}, M. and {Shah}, P. and {Sullivan}, M. and {Toy}, M. and {Abbott}, T.~M.~C. and {Aguena}, M. and {Allam}, S. and {Alves}, O. and {Annis}, J. and {Asorey}, J. and {Avila}, S. and {Bacon}, D. and {Brooks}, D. and {Burke}, D.~L. and {Carnero Rosell}, A. and {Carollo}, D. and {Carretero}, J. and {da Costa}, L.~N. and {Castander}, F.~J. and {Desai}, S. and {Diehl}, H.~T. and {Duarte}, J. and {Everett}, S. and {Ferrero}, I. and {Flaugher}, B. and {Frieman}, J. and {Garc{\'\i}a-Bellido}, J. and {Gatti}, M. and {Gaztanaga}, E. and {Giannini}, G. and {Glazebrook}, K. and {Gonz{\'a}lez-Gait{\'a}n}, S. and {Gruendl}, R.~A. and {Gutierrez}, G. and {Hinton}, S.~R. and {Hollowood}, D.~L. and {Honscheid}, K. and {James}, D.~J. and {Kuehn}, K. and {Lahav}, O. and {Lee}, S. and {Lewis}, G.~F. and {Lin}, H. and {Marshall}, J.~L. and {Mena-Fern{\'a}ndez}, J. and {Miquel}, R. and {Myles}, J. and {Nichol}, R.~C. and {Ogando}, R.~L.~C. and {Palmese}, A. and {Pereira}, M.~E.~S. and {Pieres}, A. and {Plazas Malag{\'o}n}, A.~A. and {Porredon}, A. and {Romer}, A.~K. and {Sanchez}, E. and {Sanchez Cid}, D. and {Sevilla-Noarbe}, I. and {Suchyta}, E. and {Swanson}, M.~E.~C. and {Tarle}, G. and {Tucker}, B.~E. and {Tucker}, D.~L. and {Vikram}, V. and {Walker}, A.~R. and {Weaverdyck}, N.},
        title = "{The Dark Energy Survey Supernova Program: Light Curves and 5 Yr Data Release}",
      journal = {\apj},
         year = 2024,
        month = nov,
       volume = {975},
       number = {1},
          eid = {5},
        pages = {5},
          doi = {10.3847/1538-4357/ad739a},
archivePrefix = {arXiv},
       eprint = {2406.05046},
 primaryClass = {astro-ph.CO},
       adsurl = {https://ui.adsabs.harvard.edu/abs/2024ApJ...975....5S}
}

@ARTICLE{2025arXiv251107517P,
       author = {{Popovic}, B. and {Shah}, P. and {Kenworthy}, W.~D. and {Kessler}, R. and {Davis}, T.~M. and {Goobar}, A. and {Scolnic}, D. and {Vincenzi}, M. and {Wiseman}, P. and {Chen}, R. and {Charleton}, E. and {Acevedo}, M. and {Armstrong}, P. and {Boyd}, B.~M. and {Brout}, D. and {Camilleri}, R. and {Frieman}, J. and {Galbany}, L. and {Grayling}, M. and {Kelsey}, L. and {Rose}, B. and {S{\'a}nchez}, B. and {Lee}, J. and {M{\"o}ller}, A. and {Smith}, M. and {Sullivan}, M. and {Shiamtanis}, N. and {Alarcon}, A. and {Allam}, S.~S. and {Andrade-Oliveira}, F. and {Avila}, S. and {Bacon}, D. and {Blazek}, J. and {Bocquet}, S. and {Brooks}, D. and {Burke}, D.~L. and {Carnero Rosell}, A. and {Carretero}, J. and {Cawthon}, R. and {da Costa}, L.~N. and {da Silva Pereira}, M.~E. and {Diehl}, H.~T. and {Dodelson}, S. and {Doel}, P. and {Everett}, S. and {Frohmaier}, C. and {Garc{\'\i}a-Bellido}, J. and {Gruen}, D. and {Gutierrez}, G. and {Herner}, K. and {Hinton}, S.~R. and {Hollowood}, D.~L. and {Honscheid}, K. and {Huterer}, D. and {James}, D.~J. and {Jeffrey}, N. and {Kuehn}, K. and {Lahav}, O. and {Lee}, S. and {Lidman}, C. and {Marshall}, J.~L. and {Mena-Fern{\'a}ndez}, J. and {Menanteau}, F. and {Miquel}, R. and {Muir}, J. and {Myles}, J. and {Ogando}, R.~L.~C. and {Paterno}, M. and {Plazas Malag{\'o}n}, A.~A. and {Porredon}, A. and {Prat}, J. and {Nichol}, R.~C. and {Romer}, A.~K. and {Roodman}, A. and {Sanchez}, E. and {Sanchez Cid}, D. and {Sevilla-Noarbe}, I. and {Suchyta}, E. and {Swanson}, M.~E.~C. and {To}, C. and {Tucker}, D.~L. and {Walker}, A.~R. and {Weaverdyck}, N.},
        title = "{The Dark Energy Survey Supernova Program: A Reanalysis Of Cosmology Results And Evidence For Evolving Dark Energy With An Updated Type Ia Supernova Calibration}",
      journal = {arXiv e-prints},
         year = 2025,
        month = nov,
          eid = {arXiv:2511.07517},
        pages = {arXiv:2511.07517},
          doi = {10.48550/arXiv.2511.07517},
archivePrefix = {arXiv},
       eprint = {2511.07517},
 primaryClass = {astro-ph.CO},
       adsurl = {https://ui.adsabs.harvard.edu/abs/2025arXiv251107517P}
}

@ARTICLE{2025arXiv251219783S,
       author = {{Siebert}, M.~R. and {Pierel}, J.~D.~R. and {Engesser}, M. and {Coulter}, D.~A. and {Decoursey}, C. and {Fox}, O.~D. and {Rest}, A. and {Chen}, W. and {Derkacy}, J.~M. and {Egami}, E. and {Foley}, R.~J. and {Jones}, D.~O. and {Koekemoer}, A.~M. and {Larison}, C. and {Leonard}, D.~C. and {Moriya}, T.~J. and {Quimby}, R.~M. and {Shukawa}, K. and {Strolger}, L.~G. and {Zenati}, Yossef},
        title = "{SN 2025ogs: A Spectroscopically-Normal Type Ia Supernova at z = 2 as a Benchmark for Redshift Evolution}",
      journal = {arXiv e-prints},
         year = 2025,
        month = dec,
          eid = {arXiv:2512.19783},
        pages = {arXiv:2512.19783},
archivePrefix = {arXiv},
       eprint = {2512.19783},
 primaryClass = {astro-ph.CO},
       adsurl = {https://ui.adsabs.harvard.edu/abs/2025arXiv251219783S}
}

@ARTICLE{2025EPJC...85..298O,
       author = {{Odintsov}, Sergei D. and {S{\'a}ez-Chill{\'o}n G{\'o}mez}, Diego and {Sharov}, German S.},
        title = "{Modified gravity/dynamical dark energy vs {\ensuremath{\Lambda}}CDM: is the game over?}",
      journal = {European Physical Journal C},
         year = 2025,
        month = mar,
       volume = {85},
       number = {3},
          eid = {298},
        pages = {298},
          doi = {10.1140/epjc/s10052-025-14013-3},
       adsurl = {https://ui.adsabs.harvard.edu/abs/2025EPJC...85..298O}
}

@ARTICLE{2016RPPh...79d6902K,
       author = {{Koyama}, Kazuya},
        title = "{Cosmological tests of modified gravity}",
      journal = {Reports on Progress in Physics},
         year = 2016,
        month = apr,
       volume = {79},
       number = {4},
          eid = {046902},
        pages = {046902},
          doi = {10.1088/0034-4885/79/4/046902},
archivePrefix = {arXiv},
       eprint = {1504.04623},
 primaryClass = {astro-ph.CO},
       adsurl = {https://ui.adsabs.harvard.edu/abs/2016RPPh...79d6902K}
}

@ARTICLE{2009PhRvD..80l3512W,
       author = {{Wiltshire}, David L.},
        title = "{Average observational quantities in the timescape cosmology}",
      journal = {\prd},
         year = 2009,
        month = dec,
       volume = {80},
       number = {12},
          eid = {123512},
        pages = {123512},
          doi = {10.1103/PhysRevD.80.123512},
archivePrefix = {arXiv},
       eprint = {0909.0749},
 primaryClass = {astro-ph.CO},
       adsurl = {https://ui.adsabs.harvard.edu/abs/2009PhRvD..80l3512W}
}

@ARTICLE{2011ApJS..192....1C,
       author = {{Conley}, A. and {Guy}, J. and {Sullivan}, M. and {Regnault}, N. and {Astier}, P. and {Balland}, C. and {Basa}, S. and {Carlberg}, R.~G. and {Fouchez}, D. and {Hardin}, D. and {Hook}, I.~M. and {Howell}, D.~A. and {Pain}, R. and {Palanque-Delabrouille}, N. and {Perrett}, K.~M. and {Pritchet}, C.~J. and {Rich}, J. and {Ruhlmann-Kleider}, V. and {Balam}, D. and {Baumont}, S. and {Ellis}, R.~S. and {Fabbro}, S. and {Fakhouri}, H.~K. and {Fourmanoit}, N. and {Gonz{\'a}lez-Gait{\'a}n}, S. and {Graham}, M.~L. and {Hudson}, M.~J. and {Hsiao}, E. and {Kronborg}, T. and {Lidman}, C. and {Mourao}, A.~M. and {Neill}, J.~D. and {Perlmutter}, S. and {Ripoche}, P. and {Suzuki}, N. and {Walker}, E.~S.},
        title = "{Supernova Constraints and Systematic Uncertainties from the First Three Years of the Supernova Legacy Survey}",
      journal = {\apjs},
         year = 2011,
        month = jan,
       volume = {192},
       number = {1},
          eid = {1},
        pages = {1},
          doi = {10.1088/0067-0049/192/1/1},
archivePrefix = {arXiv},
       eprint = {1104.1443},
 primaryClass = {astro-ph.CO},
       adsurl = {https://ui.adsabs.harvard.edu/abs/2011ApJS..192....1C}
}

@ARTICLE{2024ApJS..275...21G,
       author = {{Gris}, Philippe and {Awan}, Humna and {Becker}, Matthew R. and {Lin}, Huan and {Gawiser}, Eric and {Jha}, Saurabh W. and {The LSST Dark Energy Science Collaboration}},
        title = "{A Cohesive Deep Drilling Field Strategy for LSST Cosmology}",
      journal = {\apjs},
         year = 2024,
        month = dec,
       volume = {275},
       number = {2},
          eid = {21},
        pages = {21},
          doi = {10.3847/1538-4365/ad79f5},
archivePrefix = {arXiv},
       eprint = {2405.10781},
 primaryClass = {astro-ph.CO},
       adsurl = {https://ui.adsabs.harvard.edu/abs/2024ApJS..275...21G}
}

@ARTICLE{2025RSPTA.38340022E,
       author = {{Efstathiou}, George},
        title = "{Challenges to the {\ensuremath{\Lambda}} CDM cosmology}",
      journal = {Philosophical Transactions of the Royal Society of London Series A},
         year = 2025,
        month = feb,
       volume = {383},
       number = {2290},
          eid = {20240022},
        pages = {20240022},
          doi = {10.1098/rsta.2024.0022},
archivePrefix = {arXiv},
       eprint = {2406.12106},
 primaryClass = {astro-ph.CO},
       adsurl = {https://ui.adsabs.harvard.edu/abs/2025RSPTA.38340022E}
}

@ARTICLE{1992StaSc...7..457G,
       author = {{Gelman}, Andrew and {Rubin}, Donald B.},
        title = "{Inference from Iterative Simulation Using Multiple Sequences}",
      journal = {Statistical Science},
         year = 1992,
        month = jan,
       volume = {7},
        pages = {457-472},
          doi = {10.1214/ss/1177011136},
       adsurl = {https://ui.adsabs.harvard.edu/abs/1992StaSc...7..457G}
}

@article{kass1995,
   author = {Kass, Robert E and Raftery, Adrian E},
    title = {Bayes factors},
  journal = {Journal of the american statistical association},
     year = {1995},
   volume = {90},
   number = {430},
    pages = {773--795},
      doi = {10.1080/01621459.1995.10476572}
}

@inbook{Burnham_Anderson2002,
   author = {Kenneth P. Burnham and David R. Anderson},
    title = "{Model selection and multimodel inference: a practical information-theoretic approach}",
publisher = "Springer-Verlag",
  address = "New York",
     year = 2002,
}

@ARTICLE{Wilks1938,
       author = {{Wilks}, S. S.},
        title = "{The Large-Sample Distribution of the Likelihood Ratio for Testing Composite Hypotheses}",
      journal = {Annals of Mathematical Statistics},
         year = 1938,
       volume = {9},
        pages = {60-62},
          doi = {10.1214/AOMS/1177732360}
}

@ARTICLE{2020AA...641A...6P,
       author = {{Planck Collaboration} and {Aghanim}, N. and {Akrami}, Y. and {Ashdown}, M. and {Aumont}, J. and {Baccigalupi}, C. and {Ballardini}, M. and {Banday}, A.~J. and {Barreiro}, R.~B. and {Bartolo}, N. and {Basak}, S. and {Battye}, R. and {Benabed}, K. and {Bernard}, J. -P. and {Bersanelli}, M. and {Bielewicz}, P. and {Bock}, J.~J. and {Bond}, J.~R. and {Borrill}, J. and {Bouchet}, F.~R. and {Boulanger}, F. and {Bucher}, M. and {Burigana}, C. and {Butler}, R.~C. and {Calabrese}, E. and {Cardoso}, J. -F. and {Carron}, J. and {Challinor}, A. and {Chiang}, H.~C. and {Chluba}, J. and {Colombo}, L.~P.~L. and {Combet}, C. and {Contreras}, D. and {Crill}, B.~P. and {Cuttaia}, F. and {de Bernardis}, P. and {de Zotti}, G. and {Delabrouille}, J. and {Delouis}, J. -M. and {Di Valentino}, E. and {Diego}, J.~M. and {Dor{\'e}}, O. and {Douspis}, M. and {Ducout}, A. and {Dupac}, X. and {Dusini}, S. and {Efstathiou}, G. and {Elsner}, F. and {En{\ss}lin}, T.~A. and {Eriksen}, H.~K. and {Fantaye}, Y. and {Farhang}, M. and {Fergusson}, J. and {Fernandez-Cobos}, R. and {Finelli}, F. and {Forastieri}, F. and {Frailis}, M. and {Fraisse}, A.~A. and {Franceschi}, E. and {Frolov}, A. and {Galeotta}, S. and {Galli}, S. and {Ganga}, K. and {G{\'e}nova-Santos}, R.~T. and {Gerbino}, M. and {Ghosh}, T. and {Gonz{\'a}lez-Nuevo}, J. and {G{\'o}rski}, K.~M. and {Gratton}, S. and {Gruppuso}, A. and {Gudmundsson}, J.~E. and {Hamann}, J. and {Handley}, W. and {Hansen}, F.~K. and {Herranz}, D. and {Hildebrandt}, S.~R. and {Hivon}, E. and {Huang}, Z. and {Jaffe}, A.~H. and {Jones}, W.~C. and {Karakci}, A. and {Keih{\"a}nen}, E. and {Keskitalo}, R. and {Kiiveri}, K. and {Kim}, J. and {Kisner}, T.~S. and {Knox}, L. and {Krachmalnicoff}, N. and {Kunz}, M. and {Kurki-Suonio}, H. and {Lagache}, G. and {Lamarre}, J. -M. and {Lasenby}, A. and {Lattanzi}, M. and {Lawrence}, C.~R. and {Le Jeune}, M. and {Lemos}, P. and {Lesgourgues}, J. and {Levrier}, F. and {Lewis}, A. and {Liguori}, M. and {Lilje}, P.~B. and {Lilley}, M. and {Lindholm}, V. and {L{\'o}pez-Caniego}, M. and {Lubin}, P.~M. and {Ma}, Y. -Z. and {Mac{\'\i}as-P{\'e}rez}, J.~F. and {Maggio}, G. and {Maino}, D. and {Mandolesi}, N. and {Mangilli}, A. and {Marcos-Caballero}, A. and {Maris}, M. and {Martin}, P.~G. and {Martinelli}, M. and {Mart{\'\i}nez-Gonz{\'a}lez}, E. and {Matarrese}, S. and {Mauri}, N. and {McEwen}, J.~D. and {Meinhold}, P.~R. and {Melchiorri}, A. and {Mennella}, A. and {Migliaccio}, M. and {Millea}, M. and {Mitra}, S. and {Miville-Desch{\^e}nes}, M. -A. and {Molinari}, D. and {Montier}, L. and {Morgante}, G. and {Moss}, A. and {Natoli}, P. and {N{\o}rgaard-Nielsen}, H.~U. and {Pagano}, L. and {Paoletti}, D. and {Partridge}, B. and {Patanchon}, G. and {Peiris}, H.~V. and {Perrotta}, F. and {Pettorino}, V. and {Piacentini}, F. and {Polastri}, L. and {Polenta}, G. and {Puget}, J. -L. and {Rachen}, J.~P. and {Reinecke}, M. and {Remazeilles}, M. and {Renzi}, A. and {Rocha}, G. and {Rosset}, C. and {Roudier}, G. and {Rubi{\~n}o-Mart{\'\i}n}, J.~A. and {Ruiz-Granados}, B. and {Salvati}, L. and {Sandri}, M. and {Savelainen}, M. and {Scott}, D. and {Shellard}, E.~P.~S. and {Sirignano}, C. and {Sirri}, G. and {Spencer}, L.~D. and {Sunyaev}, R. and {Suur-Uski}, A. -S. and {Tauber}, J.~A. and {Tavagnacco}, D. and {Tenti}, M. and {Toffolatti}, L. and {Tomasi}, M. and {Trombetti}, T. and {Valenziano}, L. and {Valiviita}, J. and {Van Tent}, B. and {Vibert}, L. and {Vielva}, P. and {Villa}, F. and {Vittorio}, N. and {Wandelt}, B.~D. and {Wehus}, I.~K. and {White}, M. and {White}, S.~D.~M. and {Zacchei}, A. and {Zonca}, A.},
        title = "{Planck 2018 results. VI. Cosmological parameters}",
      journal = {\aap},
         year = 2020,
        month = sep,
       volume = {641},
          eid = {A6},
        pages = {A6},
          doi = {10.1051/0004-6361/201833910},
archivePrefix = {arXiv},
       eprint = {1807.06209},
 primaryClass = {astro-ph.CO},
       adsurl = {https://ui.adsabs.harvard.edu/abs/2020A&A...641A...6P}
}

\end{document}